# Linear correlations of Gibbs free energy for rare earth element oxide, hydroxide, chloride, fluoride, carbonate, and ferrite minerals and crystalline solids

Ruiguang Pan[a], Chen Zhu[a,*]

[a] Department of Earth and Atmospheric Sciences, Indiana University Bloomington, Bloomington, IN 47405, USA

*Corresponding authors:
C. Zhu: E-mail address: chenzhu@indiana.edu; ORCID ID #: 0000-0001-5374-6787

**Abstract**

Rare Earth Elements (REE) are critical minerals (metals) for the transition from fossil fuels to renewable and clean energy. Accurate thermodynamic properties of REE minerals and other crystalline solids are crucial for geochemical modeling of the solubility, speciation, and transport of REE in ore formation, extraction, chemical processing, and recycling processes. However, the Gibbs free energies of formation ($\Delta G^{o}_{f,\,\mathrm{REEX}}$) for these solids from different sources vary by 10s kJ/mol. We applied the Sverjensky linear free energy relationship (LFER) to evaluate their internal consistency and predict the unavailable $\Delta G^{o}_{f}$ of the REE solids. By considering both the effects of ionic radius size and corresponding aqueous ion properties, the Sverjensky LFER,

$$\Delta G^{o}_{f,\mathrm{REEX}} - \beta_{\mathrm{REEX}} r_{\mathrm{REE}^{Z+}} = a_{\mathrm{REEX}} \Delta G^{o}_{n,\mathrm{REE}^{Z+}} + b_{\mathrm{REEX}}$$

allows estimates with much accuracy and precision. Here, $r_{\mathrm{REE}}{}^{Z+}$ represents the Shannon-Prewitt ionic radii (Å) of $\mathrm{REE}^{Z+}$, and $\Delta G^{o}_{n,\,\mathrm{REE}}{}^{Z+}$ denotes the non-solvation contribution to the $\Delta G^{o}_{f}$ of the aqueous $\mathrm{REE}^{Z+}$ ion. X represents the remainder of the compounds. In this study, the parameters $a_{\mathrm{REEX}}$, $b_{\mathrm{REEX}}$, and $\beta_{\mathrm{REEX}}$ were regressed from $\Delta G^{o}_{\mathrm{f}}$ compilations in the literature for 13 isostructural families. Based on these linear relationships, we recommend a set of internally consistent $\Delta G^{o}_{f,\,\mathrm{REEX}}$ for 119 end-members of REE oxides, hydroxides, chlorides, fluorides, carbonates, hydrous carbonates, and ferrites. These $\Delta G^{o}_{f,\,\mathrm{REEX}}$ are combined with experimental or predicted values of $S^{o}$, $V^{o}$, and $Cp^{o}$ from the literature and incorporated into a new SUPCRT database, which allows the calculations of thermodynamic properties to high *P-T* conditions (e.g., up to 1000 $^{o}$C and 5 kb). The log $K_{\mathrm{sp}}$ of REE solid dissociation reactions were incorporated into a modified USGS program PHREEQC for calculations of speciation, solubility, and reactive transport. These thermodynamic databases will also be incorporated into the MINES database to be used together with the GEMS code package in the future.

## 1. Introduction

Rare earth elements (REE) minerals are critical minerals (metals) typically incorporated into a large variety of complex phases to provide tailored mechanical, electrical, optical, and magnetic properties. They are critical metals used for renewable energy to develop a green economy (Goodenough et al., 2018; Grandell et al., 2016). Thermodynamic properties of rare earth mineral materials control their phase stability, material compatibility, corrosion, and transformation (Migdisov et al., 2016; Navrotsky et al., 2015; Pan et al., 2024b). Therefore, accurate thermodynamic properties of those REE minerals will enhance our understanding of the chemical processes related to the formation of these REE minerals or solids, including extraction, processing, and recycling processes (Iloeje et al., 2019; Migdisov et al., 2019; Pan et al., 2024a; Zhang et al., 2016). Even though the thermodynamic properties (e.g., $\Delta G^{o}_{f}$, $\Delta H^{o}_{f}$, and $S^{o}$, symbol definitions can be found in Table 1) of REE minerals have been experimentally determined (e.g., calorimetry, etc.), the reported thermodynamic properties (e.g., Gibbs free energies of formation, $\Delta G^{o}_{f}$) of those mineral end-members from different sources vary greatly. Figure 1 shows the comparisons of standard state $\Delta G^{o}_{f}$ for cubic $Er(OH)_3$ and $CeO_2$ end-members, which vary up to 35 kJ mol$^{-1}$.

Linear correlation of $\Delta G^{o}_{f}$ values in an isostructural family of crystalline solids is a way of evaluating the internal consistency of $\Delta G^{o}_{f}$ values of the end-members. Sverjensky and Molling (1992) pioneered an empirical linear free energy correlation for crystalline solids within the same structure families. This linear correlation has been successfully applied to isostructural families including REE phosphates (Pan et al., 2024a), and carbonates, pyrochlore, zirconolite, and uranate ($MUO_4$) minerals (Wang and Xu, 1999, 2001; Xu and Wang, 1999a, b, c). The linear correlation was also used to calculate the metal partitioning between carbonate minerals and solutions (Wang and Xu, 2001) and surface precipitation constants for the sorption (Zhu, 2002). The REEs are chemically very similar to each other, showing similar crystal-chemical and solution-chemical properties (Migdisov et al., 2016). For this study, our objectives are to, first, test if the linear relationships in Sverjensky and Molling (1992) are applicable to the experimentally derived $\Delta G^{o}_{f}$ values of the trivalent and tetravalent REE minerals and, second, recommend a set of internally consistent thermodynamic database for those REE solids after evaluation of their internal consistency. Although not all REE crystalline solids reviewed in this study are minerals, we do not always say "REE minerals and other crystalline solids" to avoid cumbersomeness.

## 2. Formulation of the linear free energy correlation

In this study, we applied the linear free energy correlation for isostructural families proposed by Sverjensky and Molling (1992) to REE mineral end-members REEX, which individually share the same isostructural crystal structure with trivalent and tetravalent cations. We generally follow the procedure in Sverjensky and Molling (1992) to derive the linear correlations of the $\Delta G^{o}_{f}$ of the isostructural REE mineral families. A list of symbols and definitions is provided in Table 1. Some $\Delta G^{o}_{f}$ values of the actinide mineral end-members were used to assist in constructing the linear correlations.

The $\Delta G^{o}_{n,\mathrm{REE}^{z+}}$ non-solvation contribution to the Gibbs energy of formation ($\Delta G^{o}_{f}$) of the $\mathrm{REE}^{z+}$ ions were calculated as:

$$\Delta G^{o}_{n,\mathrm{REE}^{Z+}} = \Delta G^{o}_{f,\mathrm{REE}^{Z+}} - \Delta G^{o}_{s,\mathrm{REE}^{Z+}} \tag{1}$$

where $\Delta G^{o}_{s,\mathrm{REE}^{z+}}$ refers to the solvation contribution to the Gibbs energy of formation ($\Delta G^{o}_{f,\mathrm{REE}^{z+}}$) of the aqueous $\mathrm{REE}^{z+}$ ions. $\Delta G^{o}_{s,\mathrm{REE}^{z+}}$ can be calculated from the following equation:

$$\Delta G^{o}_{s,\mathrm{REE}^{Z+}} = \omega_{\mathrm{REE}^{Z+}} \left( {}^{1}\!/\!_{\varepsilon} - 1 \right) \tag{2}$$

where $\omega_{\mathrm{REE}^{z+}}$ denotes the conventional Born solvation coefficient for the aqueous ion $\mathrm{REE}^{z+}$ and $\varepsilon$ refers to the dielectric constant of water, which is 78.47 at 25 °C and 1 bar (Fernández et al., 1995). The value of $\omega_{\mathrm{REE}^{z+}}$ can be calculated from:

$$\omega_{\mathrm{REE}^{Z+}} = \omega^{abs}_{\mathrm{REE}^{Z+}} - Z \cdot \omega^{abs}_{\mathrm{H}^{+}} \tag{3}$$

where $\omega^{abs}_{\mathrm{H}^{+}}$ equals $2.254 \times 10^{5}$ J mol$^{-1}$ (Shock and Helgeson, 1988); $\omega^{abs}_{\mathrm{REE}^{z+}}$ refers to the absolute Born coefficient of the aqueous $\mathrm{REE}^{z+}$ ion, which can be further derived:

$$\omega^{abs}_{\mathrm{REE}^{Z+}} = (6.94657 \times 10^{5}) \cdot Z^{2} / (r_{\mathrm{e,\,REE}^{Z+}}) \tag{4}$$

where $r_{\mathrm{e,\,REE}^{z+}}$ denotes the effective electrostatic radius of the aqueous $\mathrm{REE}^{z+}$ ion, which can be obtained from the following equation:

$$r_{\mathrm{e,\,REE}^{Z+}} = r_{\mathrm{REE}^{Z+}} + Z(0.94) \tag{5}$$

where $r_{REE^{z+}}$ refers to the crystallographic radius of the aqueous $REE^{z+}$ ion, which represents its Shannon-Prewitt radius (Å) (Shannon and Prewitt, 1969) (Table 2). $Z$ in Eqs. 3–5 represents the charge of the aqueous $REE^{z+}$, which is $3^+$ for trivalent ions and $4^+$ for tetravalent ions.

The above equations were used to calculate values of $\Delta G^o_n$ (Table 2) for the metal aqueous $REE^{z+}$ cations. To present more clearly its linear correlation for $\Delta G^o_{f,\,REEX}$, $\Delta G^o_{n,\,REE^{z+}}$, and $r_{REE^{z+}}$, we expressed the equation as below and plotted the left-hand side of Eqs. (6 and 7) against the aqueous cation parameter $\Delta G^o_{n,\,REE^{z+}}$,

$$\Delta G_f^{'} = a_{REEX} \cdot \Delta G^o_{n,REE^{Z+}} + b_{REEX} \tag{6}$$

$$\Delta G_f^{'} = \Delta G^o_{f,REEX} - \beta_{REEX} \cdot r_{REE^{Z+}} \tag{7}$$

where REEX refers to the isostructural family of REE solids. $\Delta G^o_{f,\,REEX}$ refers to the standard Gibbs free energy of formation for REEX. The parameters $a_{REEX}$, $b_{REEX}$, and $\beta_{REEX}$ are regression parameters for the isostructural family of REE minerals.

## 3. Data availability

Table 2 lists the $REE^{z+}$ crystallographic radii $r$, $\Delta G^o_n$ non-solvation contributions to the $\Delta G^o_f$ of the aqueous $REE^{z+}$ ions, and $\Delta G^o_f$ of the REE end-members which are needed to carry out the regression of the free energy linear correlation.

### 3.1. $REE^{Z+}$ ion parameters (Z = 3 or 4)

REE elements occur naturally in the trivalent state but can be also in divalent (e.g., Eu) and tetravalent states (e.g., Ce, Pr, and Tb) (Atwood, 2013). Here, Z = 4 is for (Ce, Pr, and Tb)$O_2$, and Z = 3 is for all other REE minerals. The crystallographic radii of the $REE^{Z+}$ ions are needed to perform the regression of the linear free energy correlations. For REE oxides, $REE^{3+}$ ions in REE oxide have 7- (A-Type) and 6- (B- and C-Type) fold coordination numbers (Cordfunke and Konings, 2001). $REE^{4+}$ ion in $REEO_2$ oxide has 8-fold coordination numbers. For REE hydroxides, $REE^{3+}$ ions in $REE(OH)_3$ solids have 9-fold coordination numbers. $REE^{3+}$ ions in REE chlorides with $UCl_3$ and $AlCl_3$ crystal structures have 9- and 6-fold coordination numbers, respectively (Cotton, 2024). $REE^{3+}$ ions in REE fluorides have 9-fold coordination numbers (Cotton, 2024). $REE^{3+}$ ions in both carbonates ($REE_2(CO_3)_3$) and hydrous carbonates ($REE_2(CO_3)_3 \cdot 3H_2O$) have 6-fold coordination numbers. $REE^{3+}$ ions in perovskite ferrites ($REEFeO_3$) and zirconates ($REE_2Zr_2O_7$) have 12- and 8-fold coordination numbers, respectively. Detailed $REE^{Z+}$ ion parameters can be found in Table 2. The $REE^{z+}$ crystallographic radius varies with the numbers of their coordination state. In this study, the $REE^{z+}$ crystallographic radii ($r_{REE}{}^{z+}$) at different coordination numbers were taken from Shannon (1976) and a machine learning study by Baloch et al. (2021). The $\Delta G^o{}_f$ of $REE^{z+}$ aqueous ions are from Shock and Helgeson (1988), which can be found in Table 2. $\Delta G^o{}_n$ values were derived from Eqs. 1–7 and can be found in Table 2.

### 3.2. REE mineral thermodynamic parameters

REE mineral end-members show slightly different crystal structures due to their different atomic sizes and atom arrangement. For example, A-Type, B-Type, and C-Type REE oxide end-members ($REE_2O_3$) exhibit hexagonal, monoclinic, and cubic structures, respectively (Cordfunke and Konings, 2001). The molar volumes for the end-members of REE oxides were calculated from the unit cell parameters as listed in Table 2. The molar volumes of non-experimentally measured and hypothetical end-members were calculated from the correlation of the REE oxide isostructural families. The volume correlations are shown in Appendix A.

$\Delta G^o{}_f$ values at 25 °C and 1 bar of the REE oxide end-members were from the review study of Konings et al. (2014). The $\Delta G^o{}_f$ of REE hydroxides are from the recommendation of Navrotsky et al. (2015). The $\Delta G^o{}_f$ of REE chlorides are from Konings and Kovács (2003). The $\Delta G^o{}_f$ of REE fluorides are from the solubility experiments from Migdisov et al. (2009). The $\Delta G^o{}_f$ of REE carbonates are from Navrotsky et al. (2015) and Smith and Martell (1976). The $\Delta G^o{}_f$ of hydrous REE carbonates are from the calculated results from Karapet'yants et al. (1977). The $\Delta G^o{}_f$ of REE perovskite are from the recommendation of Navrotsky et al. (2015). The $\Delta G^o{}_f$ of REE zirconates are from Navrotsky et al. (2015) and Helean et al. (2001). The detailed $\Delta G^o{}_f$ of REE minerals can be found in Table 2, and their reference source can be found in the legends of the figures for the constructed linear correlations. The $\Delta G^o{}_f$ values from other studies (Konings et al., 2014; Migdisov et al., 2016; Navrotsky et al., 2015) are also included in the discussion to compare against the results retrieved or evaluated from the linear correlation study.

## 4. Results

### 4.1. REE oxides ($REE_2O_3$ and $REEO_2$)

In this study, we only cover the most common REE oxides ($REE_2O_3$ and $REEO_2$) with REE charges in 3+ and 4+. The free energies of the REE oxides within the same isostructural family were regressed using Eqs. 6 and 7, together with the values of REE ionic radii and $\Delta G^o_n$ from Table 2. The regression results for the four isostructural families are summarized in Table 3 and plotted in Fig. 2. The results show that the linear correlation lines of A-Type, B-Type, C-Type $REE_2O_3$, and cubic $REEO_2$ have $R^2$ values of 0.9995, 0.9993, 0.9970, and 0.9931, respectively, indicating excellent linear relationships for those selected thermodynamic parameters. The linear free energy relationships can be expressed as:

$$\Delta G^o_{f,A-Type} - 732.39\ r_{REE^{3+}} = 0.7422\ G^o_{n,REE^{3+}} - 2674.67 \quad (8)$$

$$\Delta G^o_{f,B-Type} - 819.51\ r_{REE^{3+}} = 1.9899\ G^o_{n,REE^{3+}} - 3120.24 \quad (9)$$

$$\Delta G^o_{f,C-Type} - 1258.34\ r_{REE^{3+}} = 2.2599\ G^o_{n,REE^{3+}} - 3647.73 \quad (10)$$

$$\Delta G^o_{f,Cubic} - 19.389\ r_{REE^{4+}} = 0.6045\ G^o_{n,REE^{4+}} - 1599.64 \quad (11)$$

where the symbols are defined in Table 1. The $\Delta G^o_f$ of the metastable forms (A-Type $Pm_2O_3$, and C-Type $Pm_2O_3$, $Sm_2O_3$, and $Eu_2O_3$) of REE oxide end-members were predicted from the correlations as well to study the phase transition thermodynamic behaviors at elevated temperatures.

The differences between experimental and calculated values of the Gibbs free energies of the formation of the solids are shown in Fig. 3. The discrepancy between calculated and experimental Gibbs free energies is less than 8 kJ $mol^{-1}$. The discrepancies are 15 kJ $mol^{-1}$ for B-Type $Pm_2O_3$, which are not used in the linear regressions. These uncertainties (<8 kJ $mol^{-1}$) are within those reported from calorimetric experiments (~<10 kJ $mol^{-1}$) (Fig. 3). The linear correlations for B- and C-Type $REE_2O_3$ are generally parallel (Fig. 2). These results indicate that the regressed linear correlations closely fit the experimentally derived $\Delta G^o_f$ of isostructural families of the four types of REE oxides.

For REE oxides, including some actinide oxides, their thermodynamic parameters have been reviewed by Konings et al. (2014). For the $REE_2O_3$ with A, B, and C-Type structures, we still recommend using the $\Delta G^o_f$ of minerals from Konings et al. (2014) and the recommended values for the predicted end-members in this study (Fig. 2) due to the excellent linear correlation (Fig. 2a). The $\Delta H^{o}_{f}$ of the $REE_2O_3$ and $REEO_2$ have been assessed by Cordfunke and Konings (2001) and directly accepted by Konings et al. (2014). To calculate the $\Delta S^{o}_{f}$ of REE oxides, we selected the $S^{o}$ of these REE metals from Konings and Beneš (2010), and the $S^{o}$ value (205.15 J $mol^{-1}$ $K^{-1}$) of $O_2$ (g) from Robie et al. (1978).

For A-Type $REE_2O_3$, the recommended $\Delta G^{o}_{f}$ of minerals shows consistent values with the previous studies (Baker and Holley, 1968; Fitzgibbon et al., 1973; Fitzgibbon et al., 1968; Huber Jr and Holley Jr, 1952; Justice and Westrum Jr, 1963; Montgomery and Hubert, 1959; Popova and Monaenkova, 1989) within the presented errors as in Fig. 2a. The reported $\Delta G^{o}_{f}$ values of minerals in Mah (1961); Schumm et al. (1973); Spedding and Miller (1952) are much higher than the recommendation. However, these values in Cordfunke and Konings (2001); Fitzgibbon et al. (1965); Hennig and Oppermann (1998); Huntelaar et al. (2000); (Kuznetsov et al., 1960); Morss et al. (1989); Oppermann et al. (1997); Putnam et al. (2000); Stubblefield et al. (1956) are significantly smaller than the recommended values in this study (Fig. 2a). For B-Type $REE_2O_3$, the recommended $\Delta G^{o}_{f}$ values of minerals are consistent with most of previous studies (Baker et al., 1972; Cordfunke and Konings, 2001; Fitzgibbon et al., 1972; Gvelesiani and Yashvili, 1967; Hennig and Oppermann, 1997; Huber Jr et al., 1964; Huber Jr et al., 1955; Schumm et al., 1973), but showing a higher value than the $\Delta G^{o}_{f}$ of $Eu_2O_3$ (B-Type) from Hennig and Oppermann (1998) and a much lower value than the $\Delta G^{o}_{f}$ of $Sm_2O_3$ (B-Type) from Spedding (1959). For C-Type $REE_2O_3$, the recommended $\Delta G^{o}_{f}$ values of minerals in this study are consistent with most of the previous studies (Huber Jr et al., 1956b, 1957; Huber Jr and Holley Jr, 1952), only showing higher $\Delta G^{o}_{f}$ values of $Y_2O_3$ (C-Type) and $Er_2O_3$ (C-Type) than Morss et al. (1993) and higher $\Delta G^{o}_{f}$ values of $Er_2O_3$ (C-Type) than Huber Jr et al. (1956a); Stuve (1965). For cubic $REEO_2$, the recommended $\Delta G^{o}_{f}$ values of minerals are consistent with the previous studies (Fitzgibbon et al., 1973; Gruber et al., 2002; Johnson et al., 1969; Justice and Westrum Jr, 1969; Morss and Konings, 2004; Schumm et al., 1973), and only showing an average $\Delta G^{o}_{f}$ value of $AmO_2$ from Eyring et al. (1952) and the recommended value from Konings et al. (2014), and a lower $\Delta G^{o}_{f}$ value of $UO_2$ than the recommended value from Konings et al. (2014).

**4.2. REE hydroxides ($REE(OH)_3$)**

REE hydroxides with $REE(OH)_3$ composition only exhibit one cubic crystal structure. The free energies of the REE hydroxides within the same isostructural family were regressed as Eq. 12, using the values of REE ionic radii and $\Delta G^{o}_{n}$ from Table 2. The regression results for this isostructural family are summarized in Table 3 and plotted in Fig. 4a.

$$\Delta G^{o}_{f,\mathrm{REE(OH)3}} + 530.0\ r_{\mathrm{REE}^{3+}} = 1.070\ G^{o}_{n,\mathrm{REE}^{3+}} - 2122.19 \quad (12)$$

In the linear correlation calculation, we didn't use the $\Delta G^{o}_{f}$ values of hydroxide-(Dy, Er, and Y) due to significant deviations from the constructed linear line (Fig. 5a). Moreover, we predicted the $\Delta G^{o}_{f}$ of hydroxide-(Pm and Lu) using the calculated linear correlation (Eq. 12). The calculated linear correlation shows an excellent $R^2$ value of 0.9944 (Table 3). The differences between experimental and calculated values of the $\Delta G^{o}_{f}$ for REE-hydroxides are shown in Fig. 4b. The discrepancy between calculated and experimental $\Delta G^{o}_{f}$ is less than 6.0 kJ $mol^{-1}$. These uncertainties are within those reported from calorimetric experiments in the literature (<10 kJ $mol^{-1}$) (Fig. 4). These results indicate that the regressed linear correlations closely fit the experimentally derived $\Delta G^{o}_{f}$ of isostructural families of REE hydroxides (Fig. 4a).

For REE hydroxides, Navrotsky et al. (2015) reviewed previously reported thermodynamic parameters, and most of the $\Delta H^{o}_{f}$ recommended by Navrotsky et al. (2015) are from Diakonov et al. (1998), as they show similar $\Delta G^{o}_{f}$-βr values in Fig. 4a. The recommended $\Delta G^{o}_{f}$ from Navrotsky et al. (2015) show close values with the study of Brookins (1988) but overall slightly smaller than the dataset from Ragavan and Adams (2009) and Rossini (1961). The $\Delta G^{o}_{f}$ of $Eu(OH)_3$ from Ragavan and Adams (2009) is extremely smaller than all others (Fig. 4a).

### 4.3. REE chlorides ($REECl_3$)

REE chlorides show two types of crystal structures, $UCl_3$ structure for (La to Gd)$Cl_3$ and $AlCl_3$ structure for (Dy to Lu and Y)$Cl_3$. $TbCl_3$ shows a unique crystal structure of $PuBr_3$ with Tb having a coordination number of 8 (Cotton, 2024). The free energies of the REE chlorides within the same isostructural family were regressed using the values of REE ionic radii and $\Delta G^{o}_{n}$ from Table 2. The linear correlations for the REE chlorides with $UCl_3$ and $AlCl_3$ crystal structures are described in Eqs. 13 and 14, respectively. The regression results for this isostructural family are summarized in Table 3 and have been plotted in Fig. 5a.

$$\Delta G^{o}_{f,\mathrm{REECl3}} + 75.01\ r_{\mathrm{REE}^{3+}} = 1.013\ G^{o}_{n,\mathrm{REE}^{3+}} - 1268.1 \quad (13)$$

$$\Delta G^{o}_{f,\mathrm{REECl3}} + 309.19\ r_{\mathrm{REE}^{3+}} = 1.017\ G^{o}_{n,\mathrm{REE}^{3+}} - 1603.73 \quad (14)$$

In the linear correlation calculation, we used all the $\Delta G^o{}_f$ of REE chlorides with $UCl_3$ structure from Konings and Kovács (2003) but didn't use the $\Delta G^o{}_f$ of $YCl_3$ from this study due to the significant deviation from the constructed linear line (Fig. 5a). Therefore, $\Delta G^o{}_f$ of $PmCl_3$ and $YCl_3$ were predicted based on these linear correlations (Eqs. 13 and 14). Because the $TbCl_3$ end-member has a unique crystal structure, which is neither $UCl_3$ nor $AlCl_3$, we directly accepted the $\Delta G^o{}_f$ value (-1010.6 kJ mol$^{-1}$) from Konings and Kovács (2003). The calculated linear lines show excellent $R^2$ values of 0.9985 and 0.9830, for REE chlorides with $UCl_3$ and $AlCl_3$ structures, respectively (Table 3). The differences between experimental and calculated values of the $\Delta G^o{}_f$ of REE-chlorides are shown in Fig. 5b. The discrepancy between calculated and experimental $\Delta G^o{}_f$ is less than 4.0 kJ mol$^{-1}$ (Fig. 5b). For REE chlorides, the constructed linear correlations for the two structures of isostructural families show excellent linear correlations using the $\Delta G^o{}_f$ values from Konings and Kovács (2003), except $YCl_3$.

#### 4.4. REE fluorides ($REEF_3$)

REE fluorides show two types of crystal structures, $LaF_3$ structure for (La to Nd)$F_3$ and $YF_3$ structure for (Sm to Lu and Y)$F_3$. The free energies of the REE fluorides within the same isostructural families from solubility experiments (Migdisov et al., 2009) were regressed (Fig. 6), using the values of REE ionic radii and $\Delta G^o{}_n$ from Table 2. The regression results for these isostructural families are summarized in Table 3 and plotted in Fig. 6a. The linear correlations for the REE fluorides with $LaF_3$ and $YF_3$ crystal structures are described by Eq. 15 and 16, respectively.

$$\Delta G^o_{f,\mathrm{REEF3}} + 78.29\, r_{\mathrm{REE^{3+}}} = 0.5665\, G^o{}_{n,\mathrm{REE^{3+}}} - 1844.46 \qquad (15)$$

$$\Delta G^o_{f,\mathrm{REEF3}} + 174.66\, r_{\mathrm{REE^{3+}}} = 0.867\, G^o{}_{n,\mathrm{REE^{3+}}} - 2021.58 \qquad (16)$$

In the linear correlation calculation, we used all the $\Delta G^o{}_f$ of REE fluorides with $LaF_3$ structure from Migdisov et al. (2016) (Fig. 6a). $\Delta G^o{}_f$ of $EuF_3$ and $YF_3$ were not reported in Migdisov et al. (2016), and therefore were predicted in this study based on Eq. 16. The calculated linear correlations show excellent $R^2$ values of 0.99750 and 0.9909, for REE fluorides with $LaF_3$ and $YF_3$ crystal structures, respectively (Table 3). The differences between experimental and calculated values of the $\Delta G^o{}_f$ of REE fluorides are shown in Fig. 6b. For all of the 13 points selected, the discrepancies between calculated and experimental $\Delta G^o{}_f$ values are less than 8.0 kJ mol$^{-1}$ for the $LaF_3$ isostructural family and less than 3 kJ mol$^{-1}$ for the $LaF_3$ isostructural family. These results indicate that the regressed linear correlations closely fit the $\Delta G^o{}_f$ derived from the solubility of REE fluorides. The $\Delta G^o{}_f$ dataset of REE fluorides from calorimetric experiments was calculated from the $\Delta H^o{}_f$ and $S^o$ of REE fluorides reported by Konings and Kovács (2003), the $S^o$ of REE metals from Konings and Beneš (2010), and the $S^o$ (205.15 J mol$^{-1}$ K$^{-1}$) of $F_2$(g) from Robie et al. (1978). This $\Delta G^o{}_f$ dataset of REE fluorides does not fall on these linear lines. The constructed correlations using the $\Delta G^o{}_f$ values from the solubility data (Migdisov et al., 2016; Migdisov et al., 2009) show much better correlations than those from Konings and Kovács (2003).

### 4.5. REE carbonates ($REE_2(CO_3)_3$) and hydrous carbonates ($REE_2(CO_3)_3 \cdot nH_2O$)

REE carbonates only show one crystal structure ($REE_2(CO_3)_3$), and hydrous REE carbonates have three crystal structures distinguished by $H_2O$ contents, as (La-Nd)$_2(CO_3)_3 \cdot 8H_2O$, (Sm-Tm, and Y)$_2(CO_3)_3 \cdot 6H_2O$, and (Yb and Lu)$_2(CO_3)_3 \cdot 3H_2O$. The free energies of the REE carbonates and hydrous carbonates within their individual isostructural family were regressed using the values of REE ionic radii and $\Delta G^o{}_n$ from Table 2. Due to the limited experiments on the reported $\Delta G^o{}_f$ of REE carbonates, we added available isostructural actinide (Am, Np, U, and Pu) carbonates to obtain a better constrained linear correlation (Fig. 7). The regression results for these isostructural families are summarized in Table 3 and plotted in Figs. 7a and 8a.

$$\Delta G^o_{f,\mathrm{REE2(CO3)3}} + 948.34\, r_{\mathrm{REE}^{3+}} = 2.098\, G^o_{n,\mathrm{REE}^{3+}} - 4640.35 \tag{17}$$

$$\Delta G^o_{f,\mathrm{REE2(CO3)3 \cdot 8H2O}} + 281.58\, r_{\mathrm{REE}^{3+}} = 1.227\, G^o_{n,\mathrm{REE}^{3+}} - 5002.35 \tag{18}$$

$$\Delta G^o_{f,\mathrm{REE2(CO3)3 \cdot 3H2O}} + 616.03\, r_{\mathrm{REE}^{3+}} = 2.0416\, G^o_{n,\mathrm{REE}^{3+}} - 4946.65 \tag{19}$$

In the linear correlation calculation, we used the reported $\Delta G^o{}_f$ of REE carbonates from Navrotsky et al. (2015) and Smith and Martell (1976), and actinide carbonates from OECD (1985). We didn't use the $\Delta G^o{}_f$ of Nd carbonate from Navrotsky et al. (2015) due to the significant deviation from the constructed linear line (Fig. 7a). Therefore, $\Delta G^o{}_f$ of non-experimentally measured REE carbonates are predicted based on the linear correlations in Eq. 17 and plotted in Fig. 7a. The calculated linear line of REE carbonate shows an excellent $R^2$ value of 0.9988 (Table 3). The differences between experimental and calculated $\Delta G^o{}_f$ values for REE carbonates are less than 6.0 kJ mol$^{-1}$ (Fig. 7b). For REE carbonates, the constructed linear correlation agrees with most of the reported $\Delta G^o{}_f$ values from (Brookins, 1983) except that the (Gd, Tm, and Lu)$_2$(CO$_3$)$_3$ show higher $\Delta G^o{}_f$ values.

For hydrous REE carbonates, we constructed the linear correlations for (La-Nd)$_2$(CO$_3$)$_3$·8H$_2$O and (Sm-Tm, and Y)$_2$(CO$_3$)$_3$·3H$_2$O, but not for (Yb and Lu)$_2$(CO$_3$)$_3$·6H$_2$O because it only has two end-members, and they not applicable to the linear correlation method. For the selected $\Delta G^o{}_f$ of (La-Nd)$_2$(CO$_3$)$_3$·8H$_2$O, we still recommend using the existing $\Delta G^o{}_f$ of (Ce-Nd)$_2$(CO$_3$)$_3$·8H$_2$O because only four end-members are selected for the linear correlations. The $\Delta G^o{}_f$ value of La$_2$(CO$_3$)$_3$·8H$_2$O was therefore predicted. In terms of (Sm-Tm, and Y)$_2$(CO$_3$)$_3$·3H$_2$O, we didn't use the $\Delta G^o{}_f$ of (Tb and Y)$_2$(CO$_3$)$_3$·3H$_2$O due to their deviations from the linear correlation, so their $\Delta G^o{}_f$ values were also predicted in this study. The differences between the retrieved values from the experimental and calculated values are shown in Fig. 8b and c. Lastly, in terms of hydrous REE carbonates (Sm-Tm, and Y)$_2$(CO$_3$)$_3$·3H$_2$O and (Ce-Nd)$_2$(CO$_3$)$_3$·8H$_2$O, the discrepancies between the calculated and experimental $\Delta G^o{}_f$ values are less than 8.0 kJ mol$^{-1}$ (Fig. 8). For REE hydrous carbonates, the constructed linear correlation for REE$_2$(CO$_3$)$_3$·3H$_2$O structure family show an excellent linear correlation when using the only available estimated mineral $\Delta G^o{}_f$ values from Karapet'yants et al. (1977). The constructed linear correlation for REE$_2$(CO$_3$)$_3$·8H$_2$O structure family is fair due to the fact that this family only has four end-members. No linear correlation of $\Delta G^o{}_f$ values for (Yb and Lu)$_2$(CO$_3$)$_3$·6H$_2$O structure family because only two end-members exist.

### 4.6. REE perovskite ferrites (REEFeO$_3$)

The free energies of the REE perovskite ferrites within the same isostructural family were regressed using the values of REE ionic radii and $\Delta G^o{}_n$ from Table 2. REEFeO$_3$ crystalline solids only exhibit one crystal structure of perovskite. The regression results for this isostructural family are summarized in Table 3 and plotted in Fig. 10a.

$$\Delta G^o_{f,\mathrm{REEFeO3}} + 450.21\, r_{\mathrm{REE}^{3+}} = 0.9150\, G^o{}_{n,\mathrm{REE}^{3+}} - 2005.97 \quad (20)$$

In the linear correlation calculation, we used all the $\Delta G^o{}_f$ of REE perovskite ferrites from Navrotsky et al. (2015) but didn't use the $\Delta G^o{}_f$ of TmFeO$_3$ from their study due to the significant deviation from the constructed linear line (Fig. 6a). We therefore predicted the $\Delta G^o{}_f$ of (Tb, Tm, Yb, Ce, and Y)FeO$_3$, but we didn't predict the $\Delta G^o{}_f$ of (Er, Lu, and Pm)FeO$_3$ due to lack of radii data of those REE ions at coordination number of 12. The calculated linear correlation shows an excellent $R^2$ value of 0.9878. The discrepancies between calculated and experimental $\Delta G^o{}_f$ values are less than 8.0 kJ mol$^{-1}$ for REE perovskite ferrites (Fig. 9b). For REE perovskite ferrites, the constructed linear correlation shows an excellent linear correlation, with only one end-member ($TmFeO_3$) deviating from the line (Fig. 9a). The thermodynamic parameters recommended by Navrotsky et al. (2015) are from Kaul et al. (1976) and Kanke and Navrotsky (1998).

### 4.7. REE zirconates ($REE_2Zr_2O_7$)

The free energies of the REE Zirconates within the same isostructural family were regressed using the values of REE ionic radii and $\Delta G^o{}_n$ from Table 2. REE zirconates with $REE_2Zr_2O_7$ formula only exhibit one crystal structure of pyrochlore. The regression results for this isostructural family are summarized in Table 3 and plotted in Fig. 10. The regressed line has a negative slope (parameter *a*) which disagrees with other isostructural families that show positive slope values of the constructed correlation. The constructed linear correlation indicates that $\Delta G'_f$ has a negative correlation with the increasing $\Delta G^o{}_n$. Therefore, here we don't recommend using the dataset of $\Delta G^o{}_f$ of REE zirconates from Navrotsky et al. (2015) and $Gd_2Zr_2O_7$ from Helean et al. (2000), as in Table 2.

## 5. Discussion

### 5.1. Comparisons between calorimetry and solubility-derived $\Delta G^o{}_f$

Here, we compare the $\Delta G^o{}_f$ values at 25 $^o$C and 1 bar for end-members within an isostructural family derived from calorimetry and solubility experiments. Both calorimetry and solubility-derived $\Delta G^o{}_f$ values generally follow a LFER trend, and it can be argued that these linear trends are parallel, but these trends are offset on the vertical axis by significant amounts. Recall that $\Delta G'_f = \Delta G^o{}_{f,\,REEX} - \beta r_{REE^{z+}}$. The volume effect of βr is the same regardless of the experimental methods for measuring $\Delta G^o{}_f$. Therefore, the difference between the calorimetric and solubility-derived $\Delta G'_f$ resulted from $\Delta G^o{}_f$.

For REE fluorides, $\Delta G^o{}_f$ from solubility data generally parallel with the calorimetric data but shows more negative values than the calorimetry data (Fig. 6). REE fluorides with the $LaF_3$ structure derived by the calorimetry data show a ~28.8 kJ/mol offset over solubility-derived $\Delta G^o{}_f$ while REE fluorites with the $YF_3$ structure has a much smaller offset, ~3.5 kJ/mol compared to solubility-derived $\Delta G^o{}_f$. In terms of the solubility products log $K_{sp}$ (e.g., $REEF_3 = REE^{3+} + 3F^-$), the offset in $\Delta G^o_f$ with $LaF_3$ structure leads to a ~5.0 units more positive log $K_{sp}$ at 25 $^o$C and 1 bar. In other words, using the $\Delta G^o_f$ values of (La, Ce, Pr, and Nd)-fluorides derived from calorimetric measurements, a geochemical model would predict that those fluorides are orders of magnitude more soluble than those using the solubility-derived $\Delta G^o_f$ values at 25 $^o$C and 1 bar in the literature. For REE fluorides with $YF_3$ structure, the small ~3.5 kJ/mol in $\Delta G^o_f$ from these two measurements is within the error of chemometric experiments (~10 kJ/mol) and only leads to a ~0.6 unit more positive log $K_{sp}$ at 25 $^o$C and 1 bar.

Similar calorimetry-solubility discrepancies in $\Delta G^o_f$ values at 25 $^o$C and 1 bar were also found for xenotime. Figure 11 shows an offset of ~22 kJ/mol between calorimetry and solubility-derived $\Delta G^o_f$ but the offset is in the opposite direction as to fluorides. The solubility-derived $\Delta G^o_f$ have more positive numbers. Using the calorimetry-derived $\Delta G^o_f$ leads to 3.9 units lower in log $K_{sp}$ for xenotime end-members, and therefore, predicting much lower xenotime solubility. On the other hand, for monazite, the calorimetry-derived $\Delta G^o_f$ values at 25 $^o$C and 1 bar agree with some solubility derived $\Delta G^o_f$ values at 25 $^o$C and 1 bar but not others (Pan et al., 2024a).

It is encouraging that both calorimetry- and solubility-derived $\Delta G^o_f$ values typically fall along a linear trend, prescribed by the Sverjensky LFER, with solubility-derived values having excellent linear correlations while calorimetry-derived values are more scattered. To resolve these discrepancies is important but this work is beyond the scope of the current study, and we will endeavor in the future to resolve them.

## 5.2. Consistency between $\Delta G^o_f$ for aqueous REE ions and REE crystalline solids

Sverjensky and Molling (1992) pointed out "A second way to use our linear free energy equation is to predict the standard Gibbs free energies of formation of aqueous cations at 25 °C and 1 bar when only the free energies of the solids are known. This is particularly useful for highly charged aqueous cations (Z = 3 or 4) ...." In other words, the LFER established in this study and Pan et al. (2024a) for 16 isostructural families provide a dataset to evaluate the internal consistency of the $\Delta G^o_f$ of aqueous REE ions as well. For the 16 linear correlations (Figs. 2-10, and those in Pan et al. (2024a)), the $\Delta G^o_f$ of aqueous REE ions ($REE^{3+}$, $Ce^{4+}$, $Pr^{4+}$, and $Tb^{4+}$) are all from Shock et al. (1997) and Shock and Helgeson (1988), which are internally consistent (Table 2). Alternative $\Delta G^o_f$ values at 25 °C and 1 bar for some REE aqueous ions have been recommended later (Gysi and Harlov, 2021; Gysi et al., 2018; Pan et al., 2024b) (Table 4). In Gysi and Harlov (2021); Gysi et al. (2018), and Pan et al. (2024b), the $\Delta G^o_f$ at 25 °C and 1 bar for some REE aqueous ions were optimized by matching with the monazite and xenotime solubility experimental data at elevated temperatures (100–250 $^{o}$C) without modifying the Helgeson-Kirkham-Flowers (HKF) parameters (Helgeson et al., 1981). In the low-temperature study of Pan et al. (2024b) the $\Delta G^{o}_{f}$ at 25 °C and 1 bar were optimized based on the rhabdophane solubility data from Gausse et al. (2016) at the temperature range of 25–90 °C.

Here, we evaluated the $\Delta G^{o}_{n}$ for aqueous $REE^{3+}$ ions calculated from the alternative $\Delta G^{o}_{f}$ and plotted them on the LFER relationships for the 16 isostructural families. The same $\Delta G^{o}_{s}$ from Eqs. 2–5 and the same $\beta$ parameters in Table 3 were used to calculate $\Delta G'_{f}$ and $\Delta G'_{n}$ values, respectively. Only experimental data used for the linear regression were plotted (Fig. 12). The $REE_2O_3$ C-Type and $REE(OH)_3$ isostructural mineral families were selected as examples (Fig. 12).

For $REE_2O_3$ C-Type, the LFER would not hold if the $\Delta G^{o}_{f}$ of $REE^{3+}$ ion recommended by Pan et al. (2024b) for high temperature experiments and by Gysi et al. (2018) and Gysi and Harlov (2021) are used. For, $REE(OH)_{3,}$ the LFER would not hold if the $\Delta G^{o}_{f}$ of $REE^{3+}$ ion recommended by Pan et al. (2024a) for high-temperature experiments and by Gysi et al. (2018) and Gysi and Harlov (2021) are used, but those from Pan et al. (2024a) that were fitted from rhabdophane solubility data at the low temperature range of 25–90 °C would hold. As noted earlier, the "alternative" $\Delta G^{o}_{f}$ at 25 °C and 1 bar for some REE aqueous ions were obtained by fitting solubility experimental data without adjusting the HKF parameters for temperature and pressure extrapolations. Solubility-derived $\Delta G^{o}_{f}$ also depends on the HKF parameters. It is beyond the scope of the current study for a systematic evaluation of HKF parameters for the REE aqueous speciation model; This will be part of the future efforts.

## 6. Conclusions

Using the linear free energy relationship (LFER) recommended by Sverjensky and Molling (1992), we evaluated the internal consistency of the Gibbs free energy values for REE minerals and other crystalline solids in literature. We found remarkable linear correlations among $\Delta G^{o}_{f}$-βr values of the experimental values against $\Delta G^{o}_{n}$ of REE ions within an isostructural family. The linear correlations also allowed us to identify the outliers and predict values for those end-members that are neither experimentally available nor appear to be outliers. Based on these evaluations, we recommend a set of standard Gibbs free energy of formation for REE minerals and crystalline solids at 25 °C and 1 bar. For completeness, we included the heat capacity, entropy, and molar volumes from the literature to enable calculations to elevated temperatures and pressures (e.g., up to 1000 °C and 5 kb) (Table 5). These recommended thermodynamic properties have been incorporated into the SUPCRTBL program (Zimmer et al., 2016), which is available at models.earth.indiana.edu. These thermodynamic databases will also be incorporated into the MINES database to be used together with the GEMS code package in the future.

## Data availability

All data generated for this study is included in the article.

## CRediT authorship contribution statement

**Ruiguang Pan:** Investigation, Data curation, Formal analysis, Visualization, Writing original draft. **Chen Zhu:** Conceptualization, Methodology, Supervision, Visualization, Writing original draft, review and editing.

## Declaration of competing interest

The authors declare that they have no known competing financial interests or personal relationships that could have appeared to influence the work reported in this paper.

## Acknowledgments

This research is based upon work partially supported by the U.S. Department of Energy, Office of Science, Office of Basic Energy Sciences, Geosciences program under Award Number DE-SC0022269 to CZ.

**Tables**

Table 1. List of symbols and definitions

| Symbol | Definition |
|---|---|
| $\Delta G^{o}_{f}$ | Gibbs free energy of formation |
| $\Delta G^{o}_{n}$ | Non-solvation contribution to the Gibbs energy of formation |
| $\Delta G^{o}_{s}$ | Solvation contribution to the Gibbs energy of formation |
| $\Delta G'_{f}$ | Adjusted Gibbs free energy of formation ($\Delta G^{o}_{f,\ \mathrm{REEX}}$ - $\beta \mathrm{r}_{\mathrm{REE}}^{z+}$) |
| $\Delta S^{o}_{f}$ | Entropy of formation from the elements |
| $S^{o}$ | Standard entropy |
| $\Delta H^{o}_{f}$ | Enthalpy of formation from the elements |
| $Cp$ | Heat capacity |
| $V^{o}$ | Molar volume |
| R | Gas constant (8.314 J $\mathrm{mol}^{-1}$ $\mathrm{K}^{-1}$) |
| $K$ | Equilibrium products |
| $K_{\mathrm{sp}}$ | Solubility constants |
| ω | Born coefficient of an ion |
| $\omega^{abs}$ | Absolute Born coefficient of an ion |
| $r_{\mathrm{REE}}^{Z+}$ | Crystallographic radius of the aqueous $\mathrm{REE}^{Z+}$ ion |
| $Z$ | Charge number of an ion |
| MX | Chemical formula of a solid |
| REE | Rare Earth Elements |

Table 2. Ionic radii and thermodynamic data for REE cations and minerals

| REE Ions | [1]C.N. | r($REE^{3+}$) Å | $\Delta G^{o}_{s}$ (kJ mol$^{-1}$) $REE^{3+}_{(aq)}$ | $\Delta G^{o}_{f}$ (kJ mol$^{-1}$) $REE^{3+}_{(aq)}$ | $\Delta G^{o}_{n}$ (kJ mol$^{-1}$) $REE^{3+}_{(aq)}$ | $V^{o}$ (cm$^{3}$ mol$^{-1}$) [2]$REE_2O_{3(s)}$ | $\Delta G^{o}_{f}$ (kJ mol$^{-1}$) [3]$REE_2O_{3(s)}$ | Crystal Structure |
|---|---|---|---|---|---|---|---|---|
| La | 7 | 1.100 | -906.99 | -686.18 | 220.82 | 49.63 | -1704.06 | A-Type |
| Ce | 7 | 1.070 | -919.13 | -676.13 | 243.00 | 47.84 | -1710.65 | A-Type |
| Pr | 7 | [4]1.056 | -924.82 | -680.32 | 244.51 | 46.68 | -1719.71 | A-Type |
| Nd | 7 | [4]1.043 | -930.43 | -671.95 | 258.48 | 45.89 | -1719.30 | A-Type |
| Pm | 6 | 0.970 | -961.00 | -665.68 | 295.33 | 45.70 | -1724.02 | B-Type |
| Sm | 6 | 0.958 | -966.17 | -665.67 | 300.50 | 45.08 | -1734.65 | B-Type |
| Eu | 6 | 0.947 | -970.94 | -574.46 | 396.48 | 44.25 | -1553.58 | B-Type |
| Gd | 6 | 0.938 | -974.87 | -663.58 | 311.29 | 43.40 | -1734.21 | B-Type |
| Cm | 6 | 0.970 | -961.00 | [5]-595.40 | 365.60 | - | -1599.82 | B-Type |
| Eu | 6 | 0.947 | -970.94 | -574.46 | 396.48 | 48.33 | -1564.72 | C-Type |
| Gd | 6 | 0.938 | -974.87 | -663.58 | 311.29 | 47.57 | -1766.57 | C-Type |
| Tb | 6 | 0.923 | -981.45 | -667.35 | 314.10 | 46.47 | -1777.84 | C-Type |
| Dy | 6 | 0.912 | -986.31 | -664.00 | 322.31 | 45.65 | -1769.92 | C-Type |
| Ho | 6 | 0.901 | -991.20 | -675.30 | 315.90 | 44.91 | -1793.34 | C-Type |
| Er | 6 | 0.890 | -996.12 | -669.02 | 327.10 | 44.16 | -1809.57 | C-Type |
| Tm | 6 | 0.880 | -1000.61 | -669.02 | 331.59 | 43.41 | -1795.52 | C-Type |
| Yb | 6 | 0.868 | -1006.04 | -640.15 | 365.89 | 42.76 | -1726.73 | C-Type |
| Lu | 6 | 0.861 | -1009.22 | -666.93 | 342.29 | 42.23 | -1792.65 | C-Type |
| Y | 6 | 0.900 | -991.65 | -685.34 | 306.31 | 44.80 | -1844.06 | C-Type |
| | | Å | $REE^{4+}_{(aq)}$ | $REE^{4+}_{(aq)}$ | $REE^{4+}_{(aq)}$ | [6]$REEO_{2(s)}$ | [3]$REEO_{2(s)}$ | |
| Ce | 8 | 0.970 | -1429.77 | -507.52 | 922.25 | 23.85 | -1027.03 | Cubic |
| Pr | 8 | 0.960 | -1434.68 | -304.18 | 1130.50 | 23.61 | -900.04 | Cubic |
| Tb | 8 | 0.880 | -1474.76 | -369.03 | 1105.73 | 21.41 | -915.41 | Cubic |
| Th | 8 | 1.050 | -1391.18 | -704.78 | 686.40 | 26.40 | -1169.24 | Cubic |
| U | 8 | 1.000 | -1415.15 | -529.90 | 885.25 | 24.64 | -1031.83 | Cubic |
| Np | 8 | 0.980 | -1424.87 | [7]-491.80 | 933.07 | 24.14 | -1026.23 | Cubic |
| Pu | 8 | 0.960 | -1434.68 | [7]-478.00 | 956.68 | 23.65 | -998.11 | Cubic |
| Am | 8 | 0.950 | -1439.62 | [8]-346.00 | 1093.62 | 23.37 | -877.03 | Cubic |
| | | Å | $REE^{3+}_{(aq)}$ | $REE^{3+}_{(aq)}$ | $REE^{3+}_{(aq)}$ | $REE(OH)_{3(s)}$ | [11]$REE(OH)_{3(s)}$ | |
| La | 9 | 1.216 | -861.74 | -686.18 | 175.56 | [9]42.79 | -1284.20 | Cubic |
| Ce | 9 | 1.196 | -869.35 | -676.13 | 193.22 | - | -1286.40 | Cubic |
| Pr | 9 | 1.179 | -875.89 | -680.32 | 195.57 | [9]40.64 | -1284.90 | Cubic |
| Nd | 9 | 1.163 | -882.09 | -671.95 | 210.14 | [9]39.44 | -1283.00 | Cubic |
| Pm | 9 | 1.144 | -889.51 | -665.68 | 223.84 | - | - | Cubic |
| Sm | 9 | 1.132 | -894.24 | -665.67 | 228.57 | [9]38.38 | -1273.30 | Cubic |
| Eu | 9 | 1.120 | -899.00 | -574.46 | 324.54 | [9]37.81 | -1180.60 | Cubic |
| Gd | 9 | 1.107 | -904.19 | -663.58 | 240.60 | [9]37.36 | -1276.20 | Cubic |
| Tb | 9 | 1.095 | -909.00 | -667.35 | 241.65 | [9]36.49 | -1281.10 | Cubic |
| Dy | 9 | 1.083 | -913.85 | -664.00 | 249.85 | [10]37.98 | -1294.80 | Cubic |
| Ho | 9 | 1.072 | -918.32 | -675.30 | 243.02 | [9]35.84 | -1297.40 | Cubic |
| Er | 9 | 1.062 | -922.40 | -669.02 | 253.38 | - | -1298.90 | Cubic |
| Tm | 9 | 1.052 | -926.51 | -669.02 | 257.49 | - | -1286.60 | Cubic |
| Yb | 9 | 1.042 | -930.64 | -640.15 | 290.49 | - | -1262.90 | Cubic |
| Lu | 9 | 1.032 | -934.79 | -666.93 | 267.86 | - | - | Cubic |
| Y | 9 | 1.075 | -917.10 | -685.34 | 231.76 | 35.96 | -1338.50 | Cubic |
| | | Å | $REE^{3+}_{(aq)}$ | $REE^{3+}_{(aq)}$ | $REE^{3+}_{(aq)}$ | [12]$REECl_{3(s)}$ | [13]$REECl_{3(s)}$ | |
| La | 9 | 1.216 | -861.74 | -686.18 | 175.56 | 63.90 | -995.89 | $UCl_3$ |
| Ce | 9 | 1.196 | -869.35 | -676.13 | 193.22 | - | -983.61 | $UCl_3$ |
| Pr | 9 | 1.179 | -875.89 | -680.32 | 195.57 | 61.40 | -982.72 | $UCl_3$ |
| Nd | 9 | 1.163 | -882.09 | -671.95 | 210.14 | 60.47 | -965.56 | $UCl_3$ |
| Sm | 9 | 1.132 | -894.24 | -665.67 | 228.57 | - | -949.55 | $UCl_3$ |
| Eu | 9 | 1.120 | -899.00 | -574.46 | 324.54 | - | -855.39 | $UCl_3$ |
| Gd | 9 | 1.107 | -904.19 | -663.58 | 240.60 | 58.03 | -943.28 | $UCl_3$ |
| Tb | 8 | 0.923 | -981.45 | -667.35 | 314.10 | - | - | $PuBr_3$ |
| Dy | 6 | 0.912 | -986.31 | -664.00 | 322.31 | 74.38 | -923.34 | $AlCl_3$ |
| Ho | 6 | 0.901 | -991.20 | -675.30 | 315.90 | - | -928.29 | $AlCl_3$ |
| Er | 6 | 0.890 | -996.12 | -669.02 | 327.10 | - | -925.02 | $AlCl_3$ |
| Tm | 6 | 0.880 | -1000.61 | -669.02 | 331.59 | - | -926.20 | $AlCl_3$ |
| Yb | 6 | 0.868 | -1006.04 | -640.15 | 365.89 | - | -892.36 | $AlCl_3$ |
| Lu | 6 | 0.861 | -1009.22 | -666.93 | 342.29 | - | -917.76 | $AlCl_3$ |
| Y | 6 | 0.900 | -991.65 | -685.34 | 306.31 | 74.83 | - | $AlCl_3$ |
| | | Å | $REE^{3+}_{(aq)}$ | $REE^{3+}_{(aq)}$ | $REE^{3+}_{(aq)}$ | [14]$REEF_{3(s)}$ | [15]$REEF_{3(s)}$ | |
| La | 9 | 1.216 | -861.74 | -686.18 | 175.56 | 33.00 | -1649.81 | $LaF_3$ |
| Ce | 9 | 1.196 | -869.35 | -676.13 | 193.22 | 32.01 | -1641.37 | $LaF_3$ |
| Pr | 9 | 1.179 | -875.89 | -680.32 | 195.57 | 31.53 | -1641.37 | $LaF_3$ |
| Nd | 9 | 1.163 | -882.09 | -671.95 | 210.14 | 30.93 | -1634.37 | $LaF_3$ |
| Sm | 9 | 1.132 | -894.24 | -665.67 | 228.57 | 31.24 | -1625.67 | $YF_3$ |
| Eu | 9 | 1.120 | -899.00 | -574.46 | 324.54 | 30.76 | - | $YF_3$ |
| Gd | 9 | 1.107 | -904.19 | -663.58 | 240.60 | 30.36 | -1619.60 | $YF_3$ |
| Tb | 9 | 1.095 | -909.00 | -667.35 | 241.65 | 29.87 | -1620.78 | $YF_3$ |
| Dy | 9 | 1.083 | -913.85 | -664.00 | 249.85 | 29.40 | -1615.77 | $YF_3$ |
| Ho | 9 | 1.072 | -918.32 | -675.30 | 243.02 | 29.02 | -1623.61 | $YF_3$ |
| Er | 9 | 1.062 | -922.40 | -669.02 | 253.38 | 28.68 | -1616.38 | $YF_3$ |
| Tm | 9 | 1.052 | -926.51 | -669.02 | 257.49 | 28.40 | -1614.56 | $YF_3$ |
| Yb | 9 | 1.042 | -930.64 | -640.15 | 290.49 | 28.16 | -1587.69 | $YF_3$ |
| Lu | 9 | 1.032 | -934.79 | -666.93 | 267.86 | 27.97 | -1609.06 | $YF_3$ |
| Y | 9 | 1.075 | -917.10 | -685.34 | 231.76 | - | - | $YF_3$ |
| | | Å | $REE^{3+}_{(aq)}$ | $REE^{3+}_{(aq)}$ | $REE^{3+}_{(aq)}$ | $REE_2(CO_3)_{3(s)}$ | $REE_2(CO_3)_{3(s)}$ | |
| La | 6 | 1.032 | -934.79 | -686.18 | 248.61 | - | [11]-3141.77 | - |
| Ce | 6 | 1.010 | -943.99 | -676.13 | 267.86 | - | - | - |
| Pr | 6 | 0.990 | -952.45 | -680.32 | 272.13 | - | - | - |
| Nd | 6 | 0.983 | -955.43 | -671.95 | 283.48 | - | [11]-3102.02 | - |
| Pm | 6 | 0.970 | -961.00 | -665.68 | 295.33 | - | - | - |
| Sm | 6 | 0.958 | -966.17 | -665.67 | 300.50 | - | [11]-3102.02 | - |
| Eu | 6 | 0.947 | -970.94 | -574.46 | 396.48 | - | - | - |
| Gd | 6 | 0.938 | -974.87 | -663.58 | 311.29 | - | - | - |
| Tb | 6 | 0.923 | -981.45 | -667.35 | 314.10 | - | - | - |
| Dy | 6 | 0.912 | -986.31 | -664.00 | 322.31 | - | [16]-3093.23 | - |
| Ho | 6 | 0.901 | -991.20 | -675.30 | 315.90 | - | - | - |
| Er | 6 | 0.890 | -996.12 | -669.02 | 327.10 | - | - | - |
| Tm | 6 | 0.880 | -1000.61 | -669.02 | 331.59 | - | - | - |
| Yb | 6 | 0.868 | -1006.04 | -640.15 | 365.89 | - | [16]-3050.55 | - |
| Lu | 6 | 0.861 | -1009.22 | -666.93 | 342.29 | - | - | - |
| Y | 6 | 0.900 | -991.65 | -685.34 | 306.31 | - | [16]-3148.04 | - |
| Am | 6 | 0.975 | -958854 | [8]-598.70 | 360.16 | - | [17]-2996.20 | - |
| Pu | 6 | 1.000 | -948210 | [7]-512.87 | 369.23 | - | [17]-2918.09 | - |
| | | Å | $REE^{3+}_{(aq)}$ | $REE^{3+}_{(aq)}$ | $REE^{3+}_{(aq)}$ | $REE_2(CO_3)_3{\cdot}nH_2O_{(s)}$ | [18]$REE_2(CO_3)_3{\cdot}nH_2O_{(s)}$ | |
| La | 6 | 1.032 | -934.79 | -686.18 | 248.61 | - | -4958.88 | - |
| Ce | 6 | 1.010 | -943.99 | -676.13 | 267.86 | - | -4958.04 | - |
| Pr | 6 | 0.990 | -952.45 | -680.32 | 272.13 | - | -4947.16 | - |
| Nd | 6 | 0.983 | -955.43 | -671.95 | 283.48 | - | -4931.26 | - |
| Pm | 6 | 0.970 | -961.00 | -665.68 | 295.33 | - | - | - |
| Sm | 6 | 0.958 | -966.17 | -665.67 | 300.50 | - | -3739.24 | - |
| Eu | 6 | 0.947 | -970.94 | -574.46 | 396.48 | - | -3553.89 | - |
| Gd | 6 | 0.938 | -974.87 | -663.58 | 311.29 | - | -3728.78 | - |
| Tb | 6 | 0.923 | -981.45 | -667.35 | 314.10 | - | -3706.19 | - |
| Dy | 6 | 0.912 | -986.31 | -664.00 | 322.31 | - | -3732.13 | - |
| Ho | 6 | 0.901 | -991.20 | -675.30 | 315.90 | - | -3748.03 | - |
| Er | 6 | 0.890 | -996.12 | -669.02 | 327.10 | - | -3736.31 | - |
| Tm | 6 | 0.880 | -1000.61 | -669.02 | 331.59 | - | -3719.58 | - |
| Y | 6 | 0.900 | -991.65 | -685.34 | 306.31 | - | -3783.59 | - |
| | | Å | $REE^{3+}_{(aq)}$ | $REE^{3+}_{(aq)}$ | $REE^{3+}_{(aq)}$ | $REEFeO_{3(s)}$ | [11]$REEFeO_{3(s)}$ | |
| La | 12 | 1.360 | -809.05 | -686.18 | 122.88 | - | -1277.80 | Perovskite |
| Pr | 12 | [4]1.304 | -829.00 | -680.32 | 148.68 | - | -1282.00 | Perovskite |
| Nd | 12 | 1.270 | -841.55 | -671.95 | 169.60 | - | -1277.40 | Perovskite |
| Sm | 12 | 1.240 | -852.70 | -665.67 | 187.02 | - | -1284.10 | Perovskite |
| Eu | 12 | [4]1.224 | -858.79 | -574.46 | 284.32 | - | -1194.50 | Perovskite |
| Gd | 12 | [4]1.212 | -863.37 | -663.58 | 199.79 | - | -1269.80 | Perovskite |
| Tb | 12 | [4]1.201 | -867.40 | -667.35 | 200.06 | - | - | Perovskite |
| Dy | 12 | [4]1.192 | -871.00 | -664.00 | 207.00 | - | -1285.30 | Perovskite |
| Ho | 12 | [4]1.183 | -874.23 | -675.30 | 198.93 | - | -1289.10 | Perovskite |
| Er | 12 | - | - | -669.02 | - | - | -1287.80 | Perovskite |
| Tm | 12 | [4]1.168 | -880.11 | -669.02 | 211.08 | - | - | Perovskite |
| Yb | 12 | [4]1.161 | -882.83 | -640.15 | 242.67 | - | - | Perovskite |
| Lu | 12 | - | - | -666.93 | - | - | - | Perovskite |
| | | Å | $REE^{3+}_{(aq)}$ | $REE^{3+}_{(aq)}$ | $REE^{3+}_{(aq)}$ | $REE_2Zr_2O_{7(s)}$ | $REE_2Zr_2O_{7(s)}$ | |
| La | 8 | 1.160 | -883.25 | -686.18 | 197.08 | - | [11]-3930.40 | Pyrochlore |
| Ce | 8 | 1.143 | -889.91 | -676.13 | 213.77 | - | [11]-3923.20 | Pyrochlore |
| Sm | 8 | 1.079 | -915.47 | -665.67 | 249.80 | - | [11]-3916.70 | Pyrochlore |
| Gd | 8 | 1.053 | -926.10 | -663.58 | 262.52 | - | [19]-3873.50 | Pyrochlore |

[1]C.N.: Coordination number.

[2]Zinkevich (2007); [3]Konings et al. (2014); [4]Baloch et al. (2021); [5]Thoenen et al. (2014); [6]Morss and Fuger (1981); [7]Lemire (2001); [8]Silva et al. (2012); [9]Mroczkowski et al. (1970); [10]Chandrasekhar et al. (2012); [11]Navrotsky et al. (2015); [12]Igarashi and Mochinaga (1987); [13]Konings and

Kovács (2003); [14]Migdisov et al. (2009); [15]Migdisov et al. (2016); [16]Martell and Smith (1974); [17]OECD (1985); [18]Karapet'yants et al. (1977); [19]Helean et al. (2000). Note that references of other thermodynamic data can be found in Section 3.2.

Table 3. Summary of regression parameters for linear free energy correlation.

| REEX | Structure | Parameters<br>a | b (kJ mol$^{-1}$) | $\beta$ (kJ Å$^{-1}$) | $R^2$ |
|---|---|---|---|---|---|
| $REE_2O_3$ | A-Type | 0.742 (-) | -2674.67 (-) | 732.39 (-) | 0.9995 |
| $REE_2O_3$ | B-Type | 2.043 (-) | -3056.77 (-) | 732.04 (-) | 0.9991 |
| $REE_2O_3$ | C-Type | 2.260 (0.178) | -3647.73 (160.52) | 1258.34 (120.38) | 0.9970 |
| $REEO_2$ | Cubic | 0.604 (0.057) | -1599.64 (207.16) | 19.39 (163.56) | 0.9930 |
| $REE(OH)_3$ | Cubic | 1.070 (0.040) | -2122.19 (42.27) | 530.11(31.50) | 0.9944 |
| $REECl_3$ | $UCl_3$ | 1.013 (0.031) | -1268.09 (54.64) | 75.10 (42.84) | 0.9985 |
| $REECl_3$ | $AlCl_3$ | 1.017 (0.147) | -1603.73 (193.47) | 390.19 (171.97) | 0.9830 |
| $REEF_3$ | $LaF_3$ | 0.566 (-) | -1844.46 (-) | 78.29 (-) | 0.9750 |
| $REEF_3$ | $YF_3$ | 0.867 (0.068) | -2021.58 (62.10) | 174.66 (44.19) | 0.9909 |
| $REE_2(CO_3)_3$ | - | 2.098 (0.0789) | -4640.35 (51.24) | 948.39 (49.84) | 0.9988 |
| $REE_2(CO_3)_3 \cdot 8H_2O$ | - | 1.227 (-) | -5002.35 (-) | -281.58 (-) | 0.9935 |
| $REE_2(CO_3)_3 \cdot 3H_2O$ | - | 2.0416 (0.114) | -4946.65 (109.66) | 616.03 (134.88) | 1.0000 |
| $REEFeO_3$ | Perovskite | 0.915 (0.070) | -2005.97 (93.38) | 450.20 (69.11) | 0.9878 |

Uncertainties of 2σ are given in parentheses for each regression parameter; Some parameters have no uncertainties because of that only 4 samples are available for the regression; All values refer to 25 °C and 1 bar.

Table 4. Thermodynamic data of $\Delta G^o_f$ (kJ mol$^{-1}$) for $REE^{3+}$ ion from different sources

| $REE^{3+}$ | Shock et al. (1997) | Pan et al. (2024) High *T* | Diff. | Pan et al. (2024) Low *T* | Diff. | Gysi et al. (2018) | Diff. | Gysi and Harlov (2021) | Diff. |
|---|---|---|---|---|---|---|---|---|---|
| La | -686.18 | -683.28 | 2.90 | -684.58 | 1.60 | | | | |
| Ce | -676.13 | -674.72 | 1.41 | -674.11 | 2.02 | -673.06 | 3.07 | | |
| Pr | -680.32 | -682.14 | -1.82 | -679.24 | 1.08 | | | | |
| Nd | -671.95 | -681.77 | -9.82 | -670.11 | 1.84 | | | | |
| Sm | -665.67 | -669.33 | -3.66 | -664.12 | 1.55 | -669.94 | -4.27 | | |
| Eu | -574.46 | -578.86 | -4.40 | -568.06 | 6.40 | | | | |
| Gd | -663.58 | -672.41 | -8.83 | -662.07 | 1.51 | -672.52 | -8.93 | | |
| Tb | -667.35 | -687.54 | -20.19 | | | | | -684.09 | -16.74 |
| Dy | -664.00 | -686.85 | -22.85 | | | | | | |
| Y | -685.34 | -706.60 | -21.26 | | | | | | |
| Ho | -675.30 | -682.67 | -7.37 | | | | | -683.66 | -8.36 |
| Er | -669.02 | -695.21 | -26.19 | | | | | | |
| Tm | -669.02 | -677.78 | -8.76 | | | | | -672.49 | -3.47 |
| Yb | -640.15 | -642.97 | -2.82 | | | | | | |
| Lu | -666.93 | -665.54 | 1.39 | | | | | -665.40 | 1.53 |

Table 5. Recommended standard thermodynamic properties of REE minerals at temperature of 298.15 K and pressure of 1 bar, and selected heat capacity function, with $T$ in Kelvin. Fonts in red represent predicted values.

| REEX | Crystal Structure | $\Delta G^o_f$ kJ mol$^{-1}$ | $\Delta H^o_f$ kJ mol$^{-1}$ | $S^o$ J K$^{-1}$ mol$^{-1}$ | $V^o$ cm$^3$ mol$^{-1}$ | $Cp = a + bT + c/T^2 + d/T^{0.5}$ <br> $a$ | $b*100$ | $c$ | $d$ | Temperature range $K$ | References |
|---|---|---|---|---|---|---|---|---|---|---|---|
| $La_2O_3$ | A-Type | -1704.06 | -1791.60 | 127.32 | 49.63 | 120.6805 | 1.342414 | -1413668 | - | 298–1800 | K14 |
| $Ce_2O_3$ | A-Type | -1710.65 | -1799.80 | 148.10 | 47.84 | 113.736 | 2.84344 | -641205 | - | 298–2392 | K14 |
| $Pr_2O_3$ | A-Type | -1719.71 | -1809.90 | 152.70 | 46.68 | 121.6594 | 2.55611 | -989420 | - | 298–2310 | K14 |
| $Nd_2O_3$ | A-Type | -1719.30 | -1806.90 | 158.70 | 45.89 | 117.1079 | 2.813655 | -1258450 | - | 298–2379 | K14 |
| $Pm_2O_3$ | A-Type | - | - | - | 44.87 | 129.454 | 1.996 | - | - | 2013–2407 | K14 |
| $Pm_2O_3$ | B-Type | -1737.64 | -1824.61 | 158.00 | 45.70 | 122.9493 | 3.00141 | -1852170 | - | 298–2013 | K14 |
| $Sm_2O_3$ | B-Type | -1734.65 | -1823.00 | 150.60 | 45.08 | 129.7953 | 1.903114 | -1862270 | - | 298–2190 | K14 |
| $Eu_2O_3$ | B-Type | -1553.58 | -1650.40 | 138.60 | 44.25 | 133.3906 | 1.66443 | -1424350 | - | 298–2327 | K14 |
| $Gd_2O_3$ | B-Type | -1734.21 | -1819.70 | 157.10 | 43.40 | 114.6104 | 1.52344 | -1249170 | - | 298–2430 | K14 |
| $Cm_2O_3$ | B-Type | -1599.82 | -1684.00 | 167.00 | 45.98 | 123.532 | 0.1455 | -1348900 | - | 298–1888 | K14 |
| $Pm_2O_3$ | C-Type | -1759.73 | - | - | 49.96 | - | - | - | - | - | K14 |
| $Sm_2O_3$ | C-Type | -1763.14 | -1826.80 | 233.42 | 49.20 | 132.4358 | 1.87799 | -2408600 | - | 298–900 | K14 |
| $Eu_2O_3$ | C-Type | -1564.72 | -1662.50 | 135.40 | 48.33 | 136.2978 | 1.49877 | -1499300 | - | 298–1350 | K14 |
| $Gd_2O_3$ | C-Type | -1766.57 | -1854.00 | 150.60 | 47.57 | 114.8086 | 1.72911 | -1283970 | - | 298–2000 | K14 |
| $Cm_2O_3$ | C-Type | -1600.92 | - | - | 50.21 | - | - | - | - | - | K14 |
| $Tb_2O_3$ | C-Type | -1777.84 | -1865.20 | 159.20 | 46.47 | 120.6682 | 2.217194 | -1002610 | - | 298–1823 | K14 |
| $Dy_2O_3$ | C-Type | -1769.92 | -1863.40 | 149.80 | 45.65 | 121.2302 | 1.527609 | -845800 | - | 298–2223 | K14 |
| $Ho_2O_3$ | C-Type | -1793.34 | -1883.30 | 156.38 | 44.91 | 121.934 | 1.011623 | -886280 | - | 298–2538 | K14 |
| $Er_2O_3$ | C-Type | -1788.61 | -1879.13 | 153.13 | 44.16 | 123.2921 | 0.862245 | -1544330 | - | 298–2538 | K14 |
| $Tm_2O_3$ | C-Type | -1795.52 | -1889.30 | 139.70 | 43.41 | 128.4322 | 0.523209 | -1178910 | - | 298–2588 | K14 |
| $Yb_2O_3$ | C-Type | -1726.73 | -1814.50 | 133.10 | 42.76 | 130.6438 | 0.334628 | -1448200 | - | 298–2687 | K14 |
| $Lu_2O_3$ | C-Type | -1792.65 | -1877.00 | 126.79 | 42.23 | 122.4593 | 0.729001 | -2034140 | - | 298–2762 | K14 |
| $Y_2O_3$ | C-Type | -1823.00 | -1911.74 | 98.96 | 44.80 | 122.91 | 0.743 | -1931300 | - | 298-2000 | MK05 |
| $CeO_2$ | Cubic | -1027.03 | -1090.40 | 62.29 | 23.85 | 74.4814 | 0.583682 | -1297100 | - | 298–3083 | K14 |
| $PrO_2$ | Cubic | -900.04 | -959.10 | 80.80 | 23.61 | 72.9881 | 1.6628 | -99900 | - | 298–663 | K14 |
| $TbO_2$ | Cubic | -915.41 | -972.20 | 86.90 | 21.41 | 73.259 | 1.32023 | 1042400 | - | 298–1400 | K14 |
| $La(OH)_3$ | Cubic | -1284.20 | -1416.10 | 117.80 | 42.79 | 521.7845 | -40.255 | 709538.23 | -5081.56 | 200–350 | D98 |
| $Ce(OH)_3$ | Cubic | -1286.40 | -1418.60 | 129.40 | 41.47 | - | - | - | - | - | - |
| $Pr(OH)_3$ | Cubic | -1284.90 | -1404.00 | 131.70 | 40.64 | 520.9956 | -40.327 | 771992.59 | -5070.02 | 10–350 | D98 |
| $Nd(OH)_3$ | Cubic | -1283.00 | -1415.60 | 129.90 | 39.44 | 254.8777 | -2.25 | 366050.62 | -2324.81 | 10–350 | D98 |
| $Pm(OH)_3$ | Cubic | -1276.30 | | | 38.97 | - | - | - | - | - | - |
| $Sm(OH)_3$ | Cubic | -1273.30 | -1406.60 | 125.80 | 38.38 | - | - | - | - | - | - |
| $Eu(OH)_3$ | Cubic | -1180.60 | -1319.10 | 119.90 | 37.81 | 518.63 | -38.286 | 705759.72 | -5035.37 | 200–350 | D98 |
| $Gd(OH)_3$ | Cubic | -1276.20 | -1408.90 | 126.60 | 37.36 | 558.2502 | -44.302 | 871657.32 | -5615.29 | 200–350 | D98 |
| $Tb(OH)_3$ | Cubic | -1281.10 | -1414.80 | 128.40 | 36.49 | 558.8346 | -44.577 | 1041050 | -5623.85 | 10–350 | D98 |
| $Dy(OH)_3$ | Cubic | -1280.79 | -1414.45 | 130.30 | 37.98 | - | - | - | - | - | - |
| $Ho(OH)_3$ | Cubic | -1297.40 | -1431.10 | 130.00 | 35.84 | 568.9153 | -44.878 | 1049500 | -5771.39 | 10–350 | D98 |
| $Er(OH)_3$ | Cubic | -1288.14 | -1421.80 | 128.60 | 35.24 | - | - | - | - | - | - |
| $Tm(OH)_3$ | Cubic | -1286.60 | -1421.10 | 126.50 | 34.80 | - | - | - | - | - | - |
| $Yb(OH)_3$ | Cubic | -1262.90 | -1395.50 | 118.60 | 34.36 | - | - | - | - | - | - |
| $Lu(OH)_3$ | Cubic | -1288.54 | -1419.01 | 117.15 | 33.93 | - | - | - | - | - | - |
| $Y(OH)_3$ | Cubic | -1304.39 | -1438.26 | 99.20 | 35.96 | 575.1682 | -46.487 | 927559.26 | -5862.91 | 10–350 | D98 |
| $LaCl_3$ | $UCl_3$ | -995.89 | -1071.52 | 137.57 | 63.90 | 74.9288 | 5.1654 | 684520 | - | 1133 | KK03 |
| $CeCl_3$ | $UCl_3$ | -983.61 | -1059.02 | 151.42 | 62.52 | 90.9772 | 3.5812 | -271530 | - | 1090 | KK03 |
| $PrCl_3$ | $UCl_3$ | -982.72 | -1058.77 | 153.30 | 61.40 | 85.6511 | 3.9524 | 134650 | - | 1060 | KK03 |
| $NdCl_3$ | $UCl_3$ | -965.56 | -1041.18 | 153.43 | 60.47 | 87.2834 | 3.8586 | 40210 | - | 1032 | KK03 |
| $PmCl_3$ | $UCl_3$ | -955.47 | - | - | - | - | - | - | - | - | - |
| $SmCl_3$ | $UCl_3$ | -949.55 | -1025.32 | 150.12 | 59.12 | 95.3748 | 3.3444 | -516135 | - | 950 | KK03 |
| $EuCl_3$ | $UCl_3$ | -855.39 | -935.41 | 144.06 | 58.50 | 100.9736 | 3.0092 | -263620 | - | 894 | KK03 |
| $GdCl_3$ | $UCl_3$ | -943.28 | -1018.20 | 151.42 | 58.03 | 88.7959 | 3.1444 | -34750 | - | 875 | KK03 |
| $TbCl_3$ | $PuBr_3$ | -1010.60 | -1085.74 | 154.85 | - | 86.292 | 3.8598 | - | - | 783 | KK03 |
| $DyCl_3$ | $AlCl_3$ | -923.34 | -994.02 | 175.40 | 74.38 | -34.7111 | 8.714 | -1978460 | 2265.502 | 924 | KK03 |
| $HoCl_3$ | $AlCl_3$ | -928.29 | -997.68 | 177.10 | 74.80 | 100.382 | 0.5091 | - | - | 993 | KK03 |
| $ErCl_3$ | $AlCl_3$ | -925.02 | -994.81 | 175.10 | 75.21 | -28.255 | 8.272 | -2055060 | 2177.088 | 1049 | KK03 |
| $TmCl_3$ | $AlCl_3$ | -926.20 | -996.09 | 173.50 | 75.59 | -38.3601 | 8.67 | -2308450 | 2382.415 | 1095 | KK03 |
| $YbCl_3$ | $AlCl_3$ | -892.36 | -959.51 | 169.30 | 76.04 | -55.7145 | 9.322 | -2731060 | 2752.211 | 1138 | KK03 |
| $LuCl_3$ | $AlCl_3$ | -917.76 | -987.12 | 153.00 | 76.31 | -74.5384 | 10.439 | -3070270 | 3000.228 | 1198 | KK03 |
| $YCl_3$ | $AlCl_3$ | -940.90 | -1013.13 | 136.82 | 74.83 | 101.7423 | 0.715464 | -1051439 | - | 298–994 | P84 |
| $LaF_3$ | $LaF_3$ | -1656.37 | -1732.05 | 106.98 | - | -329.312 | 16.191 | -15085500 | 9212.611 | 1766 | KK03 |
| $CeF_3$ | $LaF_3$ | -1641.37 | -1718.50 | 115.23 | - | -59.1137 | 8.063 | -4575500 | 3075.152 | 1703 | KK03 |
| $PrF_3$ | $LaF_3$ | -1641.37 | -1717.91 | 121.22 | - | 14.00406 | 3.548 | -5381370 | 2197.807 | 1670 | KK03 |
| $NdF_3$ | $LaF_3$ | -1634.37 | -1710.64 | 120.79 | - | 37.82125 | 4.02 | -2617590 | 1231.834 | 1649 | KK03 |
| $SmF_3$ | $YF_3$ | -1624.37 | -1701.09 | 116.50 | - | 169.0564 | -7.6809 | -4840760 | - | 743 | KK03 |
| $EuF_3$ | $YF_3$ | -1544.55 | -1625.63 | 110.07 | - | 117.4275 | - | -1735890 | - | 973 | KK03 |
| $GdF_3$ | $YF_3$ | -1621.37 | -1698.14 | 114.77 | - | 102.3403 | 0.60945 | -1401620 | - | 1347 | KK03 |
| $TbF_3$ | $YF_3$ | -1618.37 | -1695.14 | 118.97 | - | 97.5769 | 1.98845 | -1156100 | - | 1446 | KK03 |
| $DyF_3$ | $YF_3$ | -1616.87 | -1695.57 | 118.07 | - | 107.776 | 1.763 | -586828.7 | -299.813 | 1426 | KK03 |
| $HoF_3$ | $YF_3$ | -1624.37 | -1701.61 | 120.34 | - | -98.1651 | 8.272 | -7160320 | 4159.99 | 1416 | KK03 |
| $ErF_3$ | $YF_3$ | -1614.37 | -1692.45 | 116.86 | - | 6.16555 | 4.455 | -4469530 | 2073.427 | 1388 | KK03 |
| $TmF_3$ | $YF_3$ | -1614.37 | -1692.63 | 114.98 | - | 53.9644 | 2.401 | -3044090 | 1097.046 | 1325 | KK03 |
| $YbF_3$ | $YF_3$ | -1587.37 | -1662.58 | 111.84 | - | 103.7012 | 0.92366 | -1516080 | - | 1267 | KK03 |
| $LuF_3$ | $YF_3$ | -1610.37 | -1688.00 | 94.83 | - | 89.0368 | 1.92857 | -685980 | - | 1230 | KK03 |
| $YF_3$ | $YF_3$ | -1632.86 | -1710.36 | 88.70 | - | 106.8418 | 0.16485 | -434000 | - | 298–1350 | P84 |
| $LaFeO_3$ | Perovskite | -1277.80 | -1351.2 | 118.00 | - | - | - | - | - | - | - |
| $CeFeO_3$ | Perovskite | -1274.59 | - | - | - | - | - | - | - | - | - |
| $PrFeO_3$ | Perovskite | -1282.00 | -1364.6 | 102.70 | 57.48 | - | - | - | - | - | - |
| $NdFeO_3$ | Perovskite | -1277.40 | -1359.9 | 116.00 | 57.26 | - | - | - | - | - | - |
| $SmFeO_3$ | Perovskite | -1284.10 | -1367 | 161.50 | | - | - | - | - | - | - |
| $EuFeO_3$ | Perovskite | -1194.50 | -1280.6 | 188.00 | - | - | - | - | - | - | - |
| $GdFeO_3$ | Perovskite | -1269.80 | -1365.3 | 129.00 | - | - | - | - | - | - | - |
| $TbFeO_3$ | Perovskite | -1282.19 | -1383.9 | 65.91 | - | - | - | - | - | - | - |
| $DyFeO_3$ | Perovskite | -1285.30 | -1378.8 | 126.80 | - | - | - | - | - | - | - |
| $HoFeO_3$ | Perovskite | -1289.10 | -1388.7 | 131.00 | - | - | - | - | - | - | - |
| $ErFeO_3$ | Perovskite | -1287.80 | -1397.1 | 137.10 | - | - | - | - | - | - | - |
| $TmFeO_3$ | Perovskite | -1286.96 | - | 86.00 | - | - | - | - | - | - | - |
| $YbFeO_3$ | Perovskite | -1261.20 | -1354.3 | 82.44 | - | - | - | - | - | - | - |
| $LuFeO_3$ | Perovskite | - | -1385.6 | - | - | - | - | - | - | - | - |
| $YFeO_3$ | Perovskite | -1296.40 | - | - | - | - | - | - | - | - | - |
| $La_2(CO_3)_3$ | - | -3141.77 | -3378.05 | 261.08 | - | - | - | - | - | - | - |
| $Ce_2(CO_3)_3$ | - | -3120.57 | -3357.69 | 284.51 | - | - | - | - | - | - | - |
| $Pr_2(CO_3)_3$ | - | -3130.57 | -3369.34 | 287.02 | - | - | - | - | - | - | - |
| $Nd_2(CO_3)_3$ | - | -3113.40 | -3349.75 | 292.46 | - | - | - | - | - | - | - |
| $Pm_2(CO_3)_3$ | - | -3100.89 | - | - | - | - | - | - | - | - | - |
| $Sm_2(CO_3)_3$ | - | -3102.02 | -3338.95 | 284.93 | - | - | - | - | - | - | - |
| $Eu_2(CO_3)_3$ | - | -2910.50 | -3153.69 | 280.33 | - | - | - | - | - | - | - |
| $Gd_2(CO_3)_3$ | - | -3097.75 | -3333.89 | 284.51 | - | - | - | - | - | - | - |
| $Tb_2(CO_3)_3$ | - | -3106.07 | -3343.95 | 287.02 | - | - | - | - | - | - | - |
| $Dy_2(CO_3)_3$ | - | -3093.23 | -3335.43 | 283.68 | - | - | - | - | - | - | - |
| $Ho_2(CO_3)_3$ | - | -3123.16 | -3361.30 | 292.04 | - | - | - | - | - | - | - |
| $Er_2(CO_3)_3$ | - | -3110.11 | -3348.97 | 288.28 | - | - | - | - | - | - | - |
| $Tm_2(CO_3)_3$ | - | -3110.16 | -3352.64 | 273.63 | - | - | - | - | - | - | - |
| $Yb_2(CO_3)_3$ | - | -3050.55 | -3287.05 | 266.94 | - | - | - | - | - | - | - |
| $Lu_2(CO_3)_3$ | - | -3105.73 | -3343.78 | 243.93 | - | - | - | - | - | - | - |
| $Y_2(CO3)_3$ | - | -3148.04 | -3385.43 | 233.05 | - | - | - | - | - | - | - |
| $La_2(CO_3)_3·8H_2O$ | - | -4958.88 | -5688.65 | 471.96 | - | - | - | - | - | - | - |
| $Ce_2(CO_3)_3·8H_2O$ | - | -4958.04 | -5670.43 | 556.47 | - | - | - | - | - | - | - |
| $Pr_2(CO_3)_3·8H_2O$ | - | -4947.16 | -5699.38 | 430.95 | - | - | - | - | - | - | - |
| $Nd_2(CO_3)_3·8H_2O$ | - | -4931.26 | - | - | - | - | - | - | - | - | - |
| $Sm_2(CO_3)_3·3H_2O$ | - | -3739.24 | -4521.43 | 322.17 | - | - | - | - | - | - | - |
| $Eu_2(CO_3)_3.3H_2O$ | - | -3553.89 | -4356.07 | 271.54 | - | - | - | - | - | - | - |
| $Gd_2(CO_3)_3·3H_2O$ | - | -3728.78 | -4510.92 | 319.24 | - | - | - | - | - | - | - |
| $Tb_2(CO_3)_3·3H_2O$ | - | -3736.78 | -4528.28 | 296.23 | - | - | - | - | - | - | - |
| $Dy_2(CO_3)_3·3H_2O$ | - | -3732.13 | -4509.73 | 353.97 | - | - | - | - | - | - | - |
| $Ho_2(CO_3)_3·3H_2O$ | - | -3748.03 | -4522.57 | 358.99 | - | - | - | - | - | - | - |
| $Er_2(CO_3)_3·3H_2O$ | - | -3736.31 | -4510.58 | 358.57 | - | - | - | - | - | - | - |
| $Tm_2(CO_3)_3·3H_2O$ | - | -3719.58 | -4483.86 | 389.53 | - | - | - | - | - | - | - |
| $Yb_2(CO_3)_3·6H_2O$ | - | -4389.02 | - | - | - | - | - | - | - | - | - |
| $Lu_2(CO_3)_3·6H_2O$ | - | -4356.38 | - | - | - | - | - | - | - | - | - |
| $Y_2(CO_3)_3 \cdot 3H_2O$ | - | -3766.87 | -4539.03 | 305.43 | - | - | - | - | - | - | - |

Abbreviations: K14: Konings et al. (2014); MK05: Morss and Konings (2004); D98: Diakonov et al. (1998); KK03: Konings and Kovács (2003); P84: Pankratz (1984).

**Figures**

Figure 1

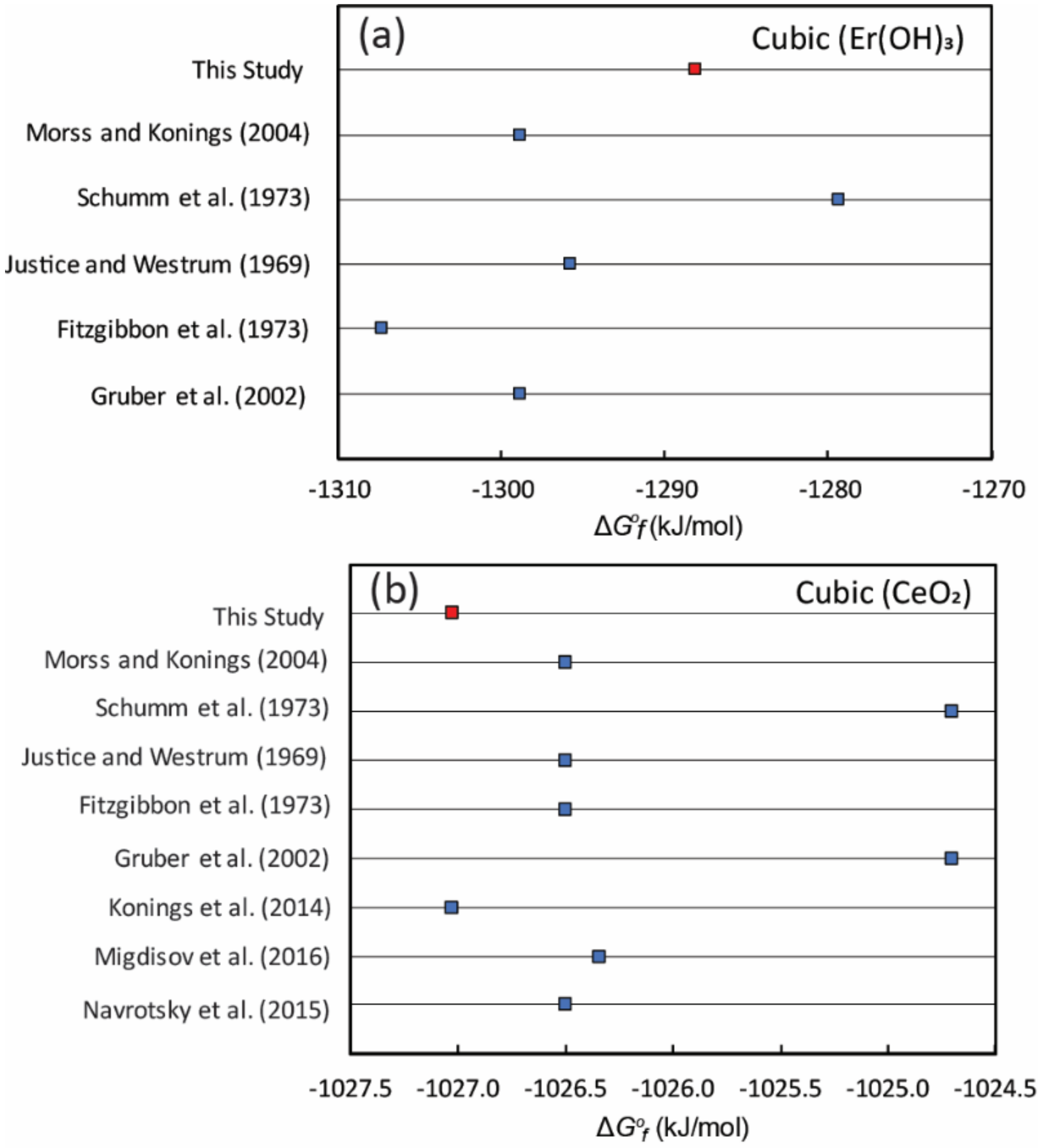

**Figure 1.** Examples of the $\Delta G^{o}_{f}$ scatter in the literature at 25 °C and 1 bar for REE crystalline solids. **(a)** Cubic $Er(OH)_3$ and **(b)** Cubic $CeO_2$ crystalline solids. All cited dark symbols are calorimetric measurements.

Figure 2

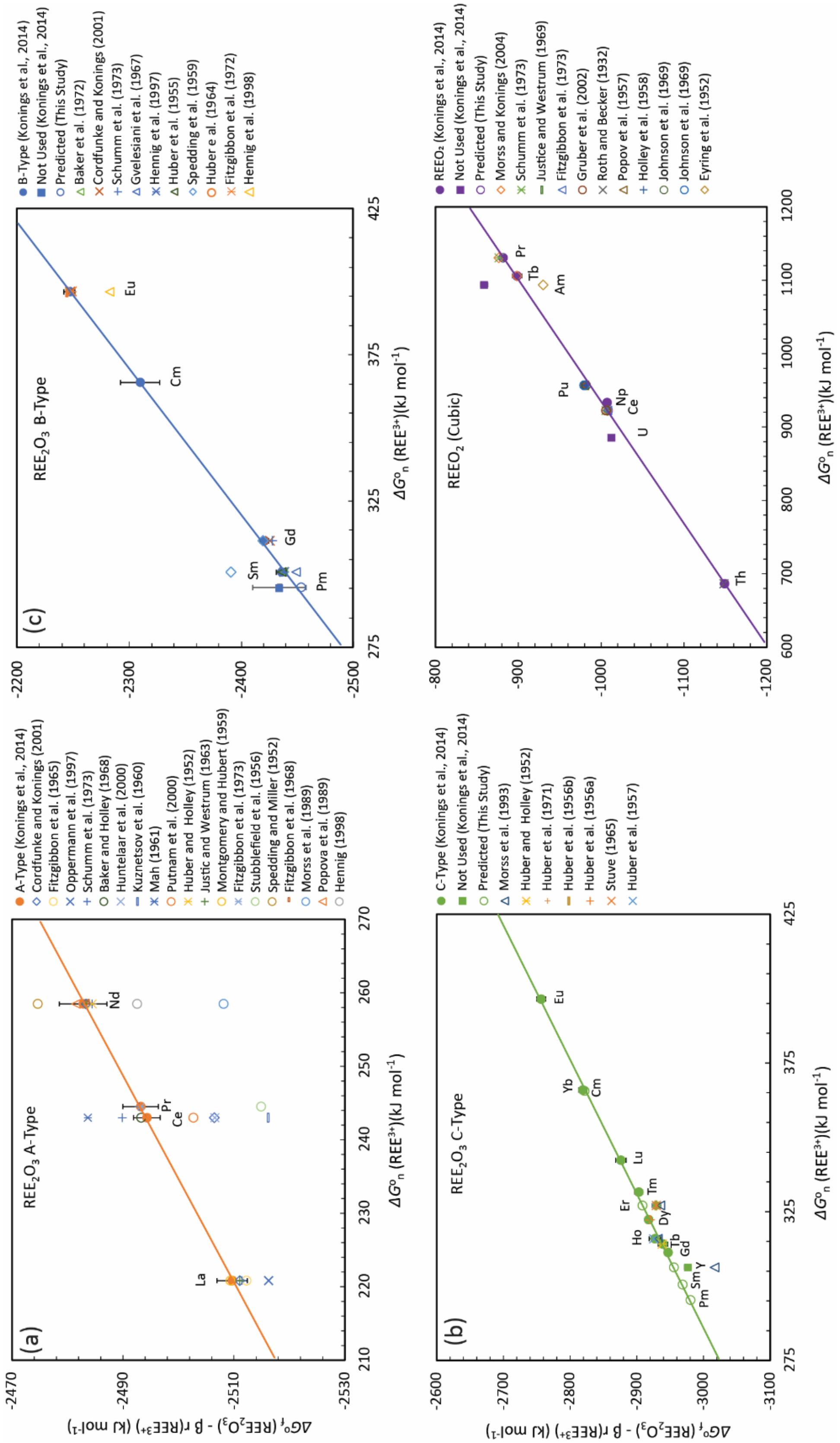
(a)
REE2O3 A-Type
ΔG°f (REE2O3) - β r(REE3+) (kJ mol-1)
ΔG°n (REE3+)(kJ mol-1)
A-Type (Konings et al., 2014)
Cordfunke and Konings (2001)
Fitzgibbon et al. (1965)
Oppermann et al. (1997)
Schumm et al. (1973)
Baker and Holley (1968)
Huntelaar et al. (2000)
Kuznetsov et al. (1960)
Mah (1961)
Putnam et al. (2000)
Huber and Holley (1952)
Justic and Westrum (1963)
Montgomery and Hubert (1959)
Fitzgibbon et al. (1973)
Stubblefield et al. (1956)
Spedding and Miller (1952)
Fitzgibbon et al. (1968)
Morss et al. (1989)
Popova et al. (1989)
Hennig (1998)
(b)
REE2O3 C-Type
C-Type (Konings et al., 2014)
Not Used (Konings et al., 2014)
Predicted (This Study)
Morss et al. (1993)
Huber and Holley (1952)
Huber et al. (1971)
Huber et al. (1956b)
Huber et al. (1956a)
Stuve (1965)
Huber et al. (1957)
(c)
REE2O3 B-Type
B-Type (Konings et al., 2014)
Not Used (Konings et al., 2014)
Predicted (This Study)
Baker et al. (1972)
Cordfunke and Konings (2001)
Schumm et al. (1973)
Gvelesiani et al. (1967)
Hennig et al. (1997)
Huber et al. (1955)
Spedding et al. (1959)
Huber e al. (1964)
Fitzgibbon et al. (1972)
Hennig et al. (1998)
REEO2 (Cubic)
REEO2 (Konings et al., 2014)
Not Used (Konings et al., 2014)
Predicted (This Study)
Morss and Konings (2004)
Schumm et al. (1973)
Justice and Westrum (1969)
Fitzgibbon et al. (1973)
Gruber et al. (2002)
Roth and Becker (1932)
Popov et al. (1957)
Holley et al. (1958)
Johnson et al. (1969)
Johnson et al. (1969)
Eyring et al. (1952)

**Figure 2.** Graphical representations of linear correlations of isostructural **(a)** A-Type $REE_2O_3$, **(b)** B-Type $REE_2O_3$, **(c)** C-Type $REE_2O_3$, and **(d)** cubic $REEO_2$ mineral families and comparisons with previous studies. Solid circles represent the experimental data used for regression. Solid squares are experimental data not used in regression. Filled circles stand for calorimetry data and other symbols of solubility data. Predicted values show as open circles.

Figure 3

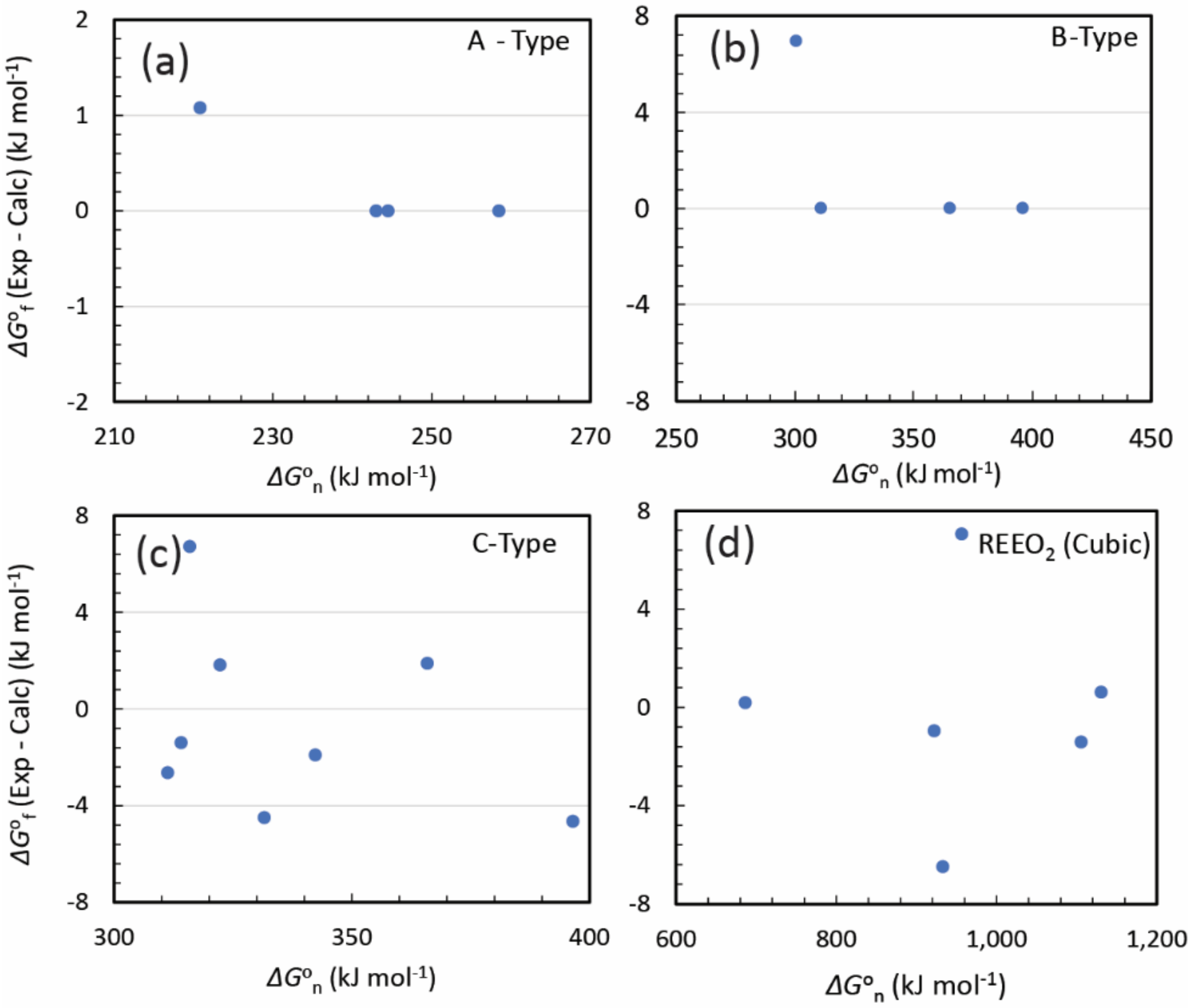

**Figure 3.** Residuals between the experimentally measured and calculated $\Delta G^{o}_{f}$ values of the crystalline solids from the four isostructural mineral families. **(a)** A-Type $REE_2O_3$, **(b)** B-Type $REE_2O_3$, **(c)** C-Type $REE_2O_3$, and **(d)** cubic $REEO_2$.

Figure 4

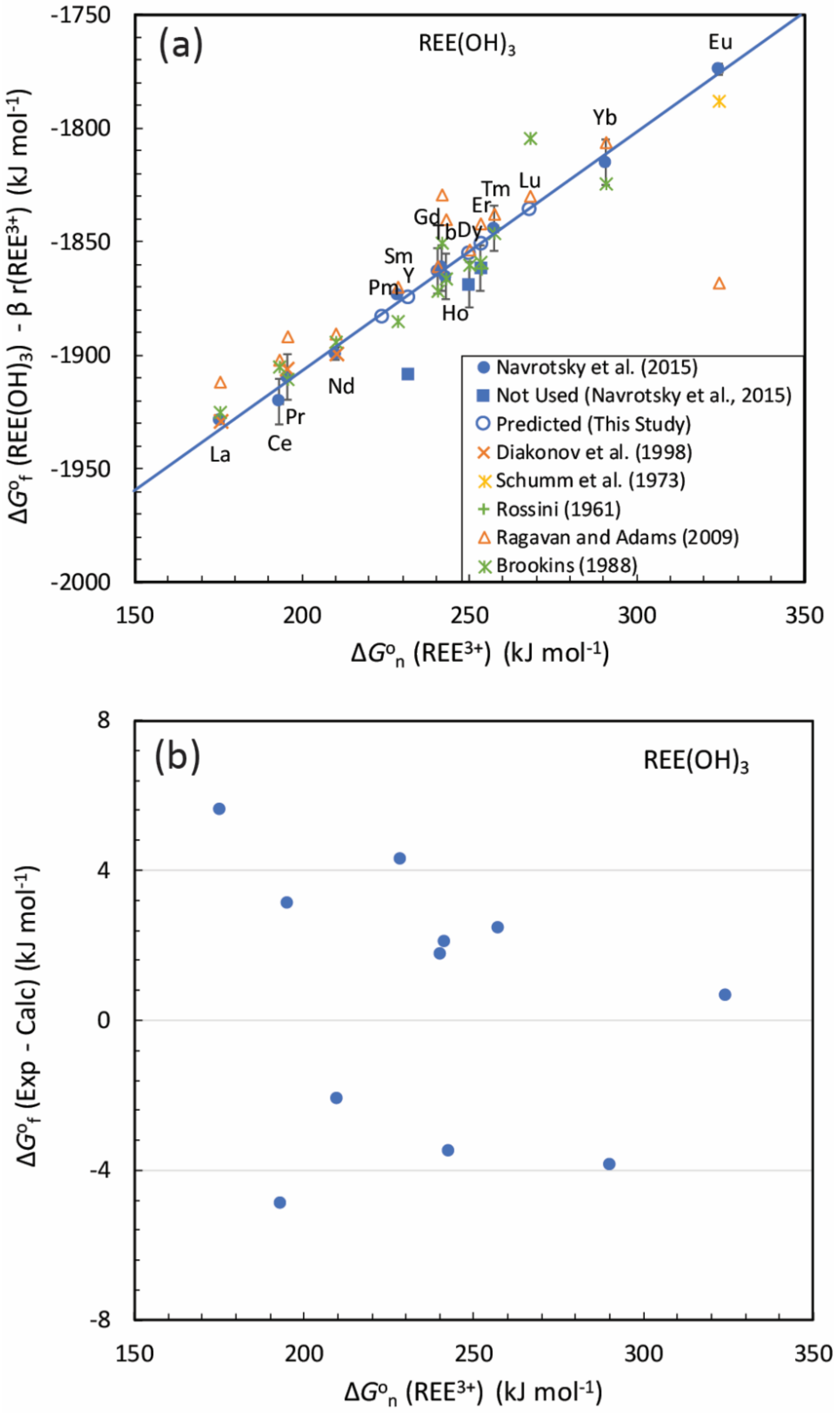

**Figure 4. (a)** Graphical representations of linear correlations of isostructural $REE(OH)_3$ mineral families and comparisons with previous studies, and **(b)** Residuals between the experimentally measured and calculated $\Delta G^o{}_f$ values for $REE(OH)_3$ isostructural family. Solid circles are used in regressions and solid squares are not used in regression. Bars represent the uncertainties of $\Delta G^o{}_f$ values.

Figure 5

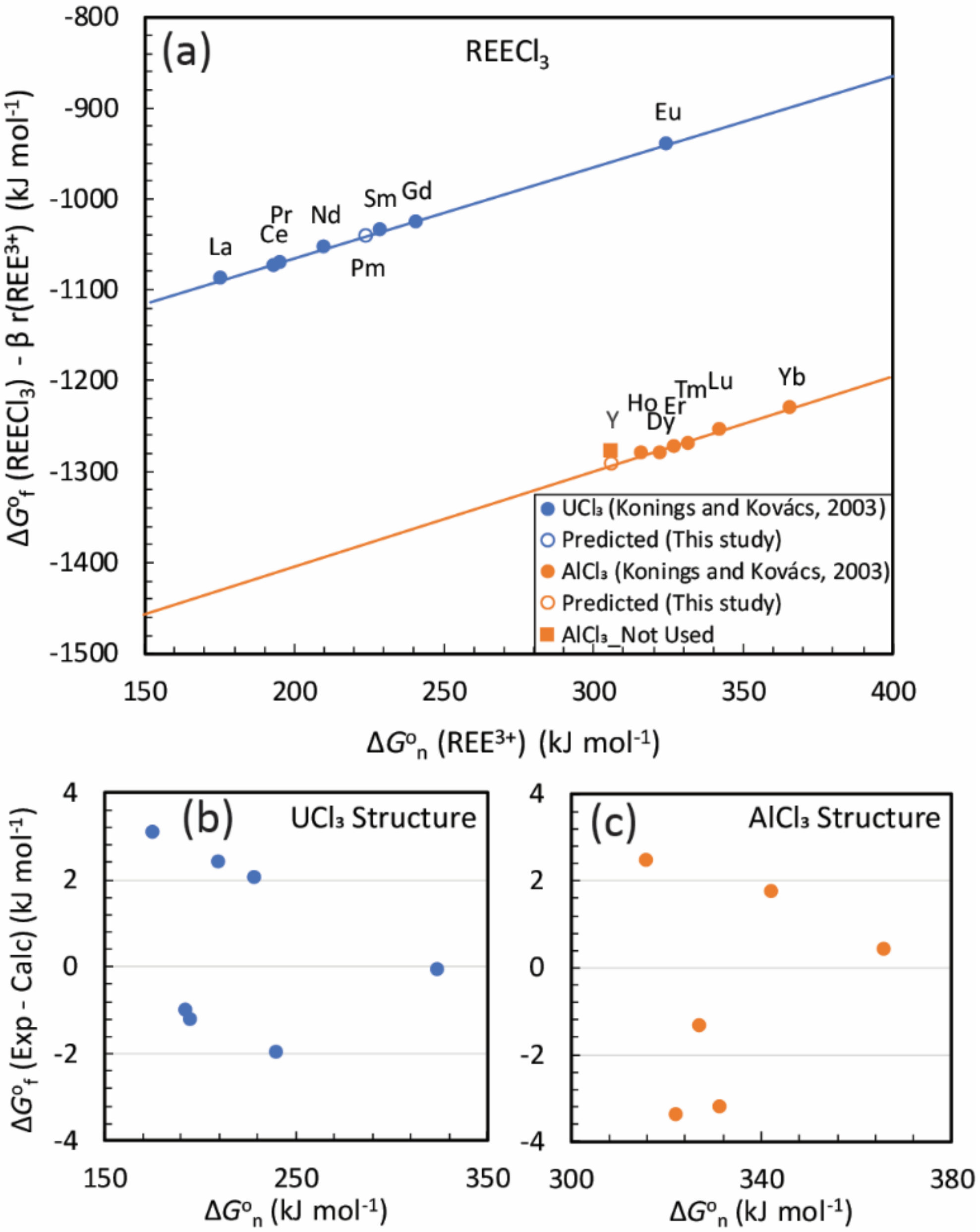

**Figure 5. (a)** Graphical representations of linear correlations of isostructural $REECl_3$ mineral families ($UCl_3$ and $AlCl_3$ structures), and **(b)** Residuals between the experimentally measured and calculated $\Delta G^o{}_f$ values for $REECl_3$ isostructural family.

Figure 6

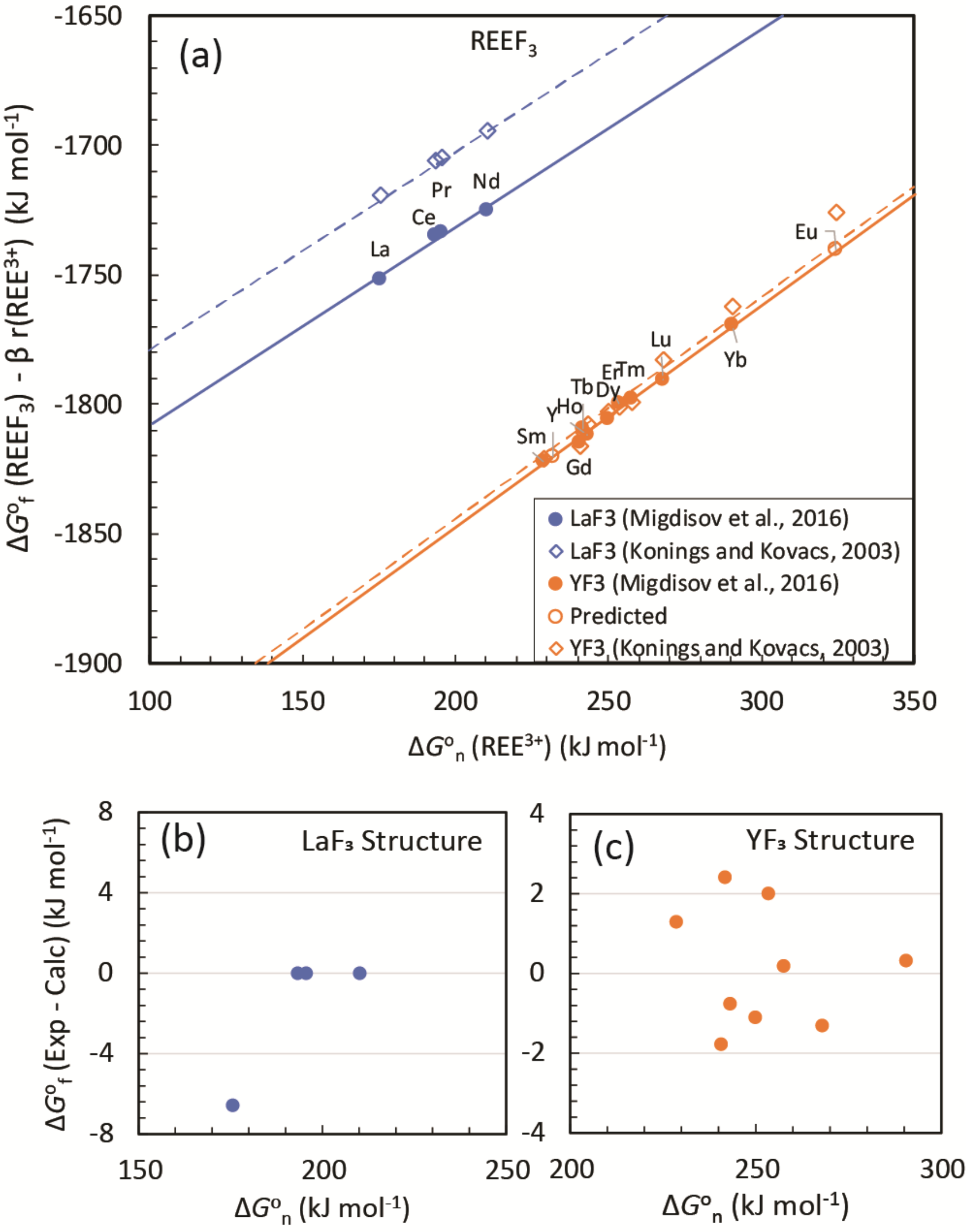

**Figure 6. (a)** Graphical representations of linear correlations of isostructural $REEF_3$ mineral families ($LaF_3$ and $YF_3$ structures) and comparisons with previous calorimetry studies Konings and Kovács (2003), and **(b-c)**. Residuals between the experimentally measured and calculated $\Delta G^o_f$ values for $REEF_3$ isostructural family. Filled circles denote the data from mineral solubility experiments from Migdisov et al. (2016), and open diamonds denote the calorimetric measurements from Konings and Kovács (2003).

Figure 7

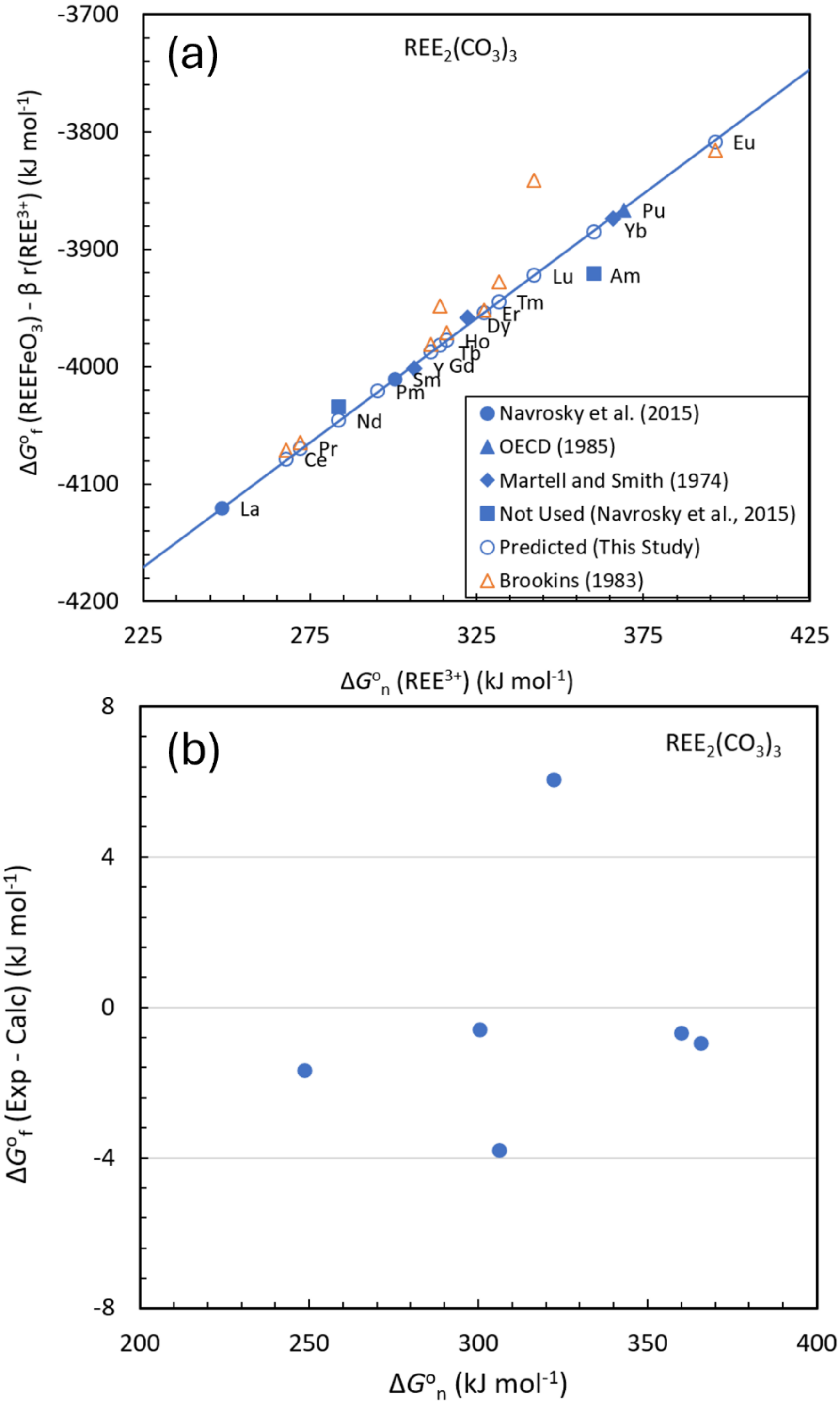

**Figure 7. (a)** Graphical representations of linear correlations of isostructural $REE_2(CO)_3$ mineral families and comparisons with previous studies (Brookins, 1988; Martell and Smith, 1974; Navrotsky et al., 2015; OECD, 1985), and **(b)** Residuals between the experimentally measured and calculated $\Delta G^o{}_f$ values for $REE_2(CO)_3$ isostructural family.

Figure 8

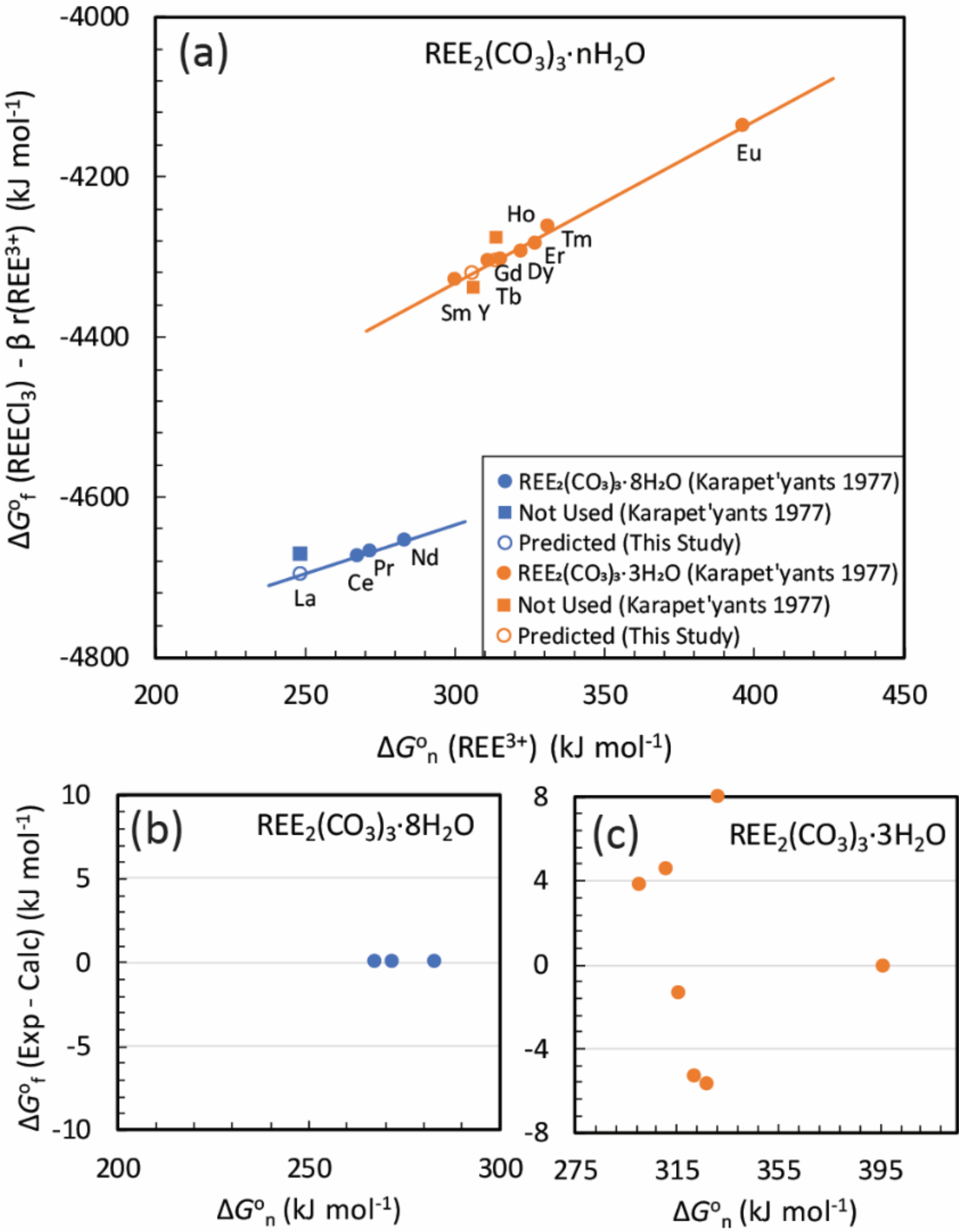

**Figure 8. (a)** Graphical representations of linear correlations of isostructural $REE_2(CO)_3 \cdot nH_2O$ mineral families, and residuals between the originally reported in Karapet'yants et al. (1977) and calculated $\Delta G^o_f$ values for **(b)** $REE_2(CO)_3 \cdot 8H_2O$ and **(c)** $REE_2(CO)_3 \cdot 3H_2O$ isostructural family.

Figure 9

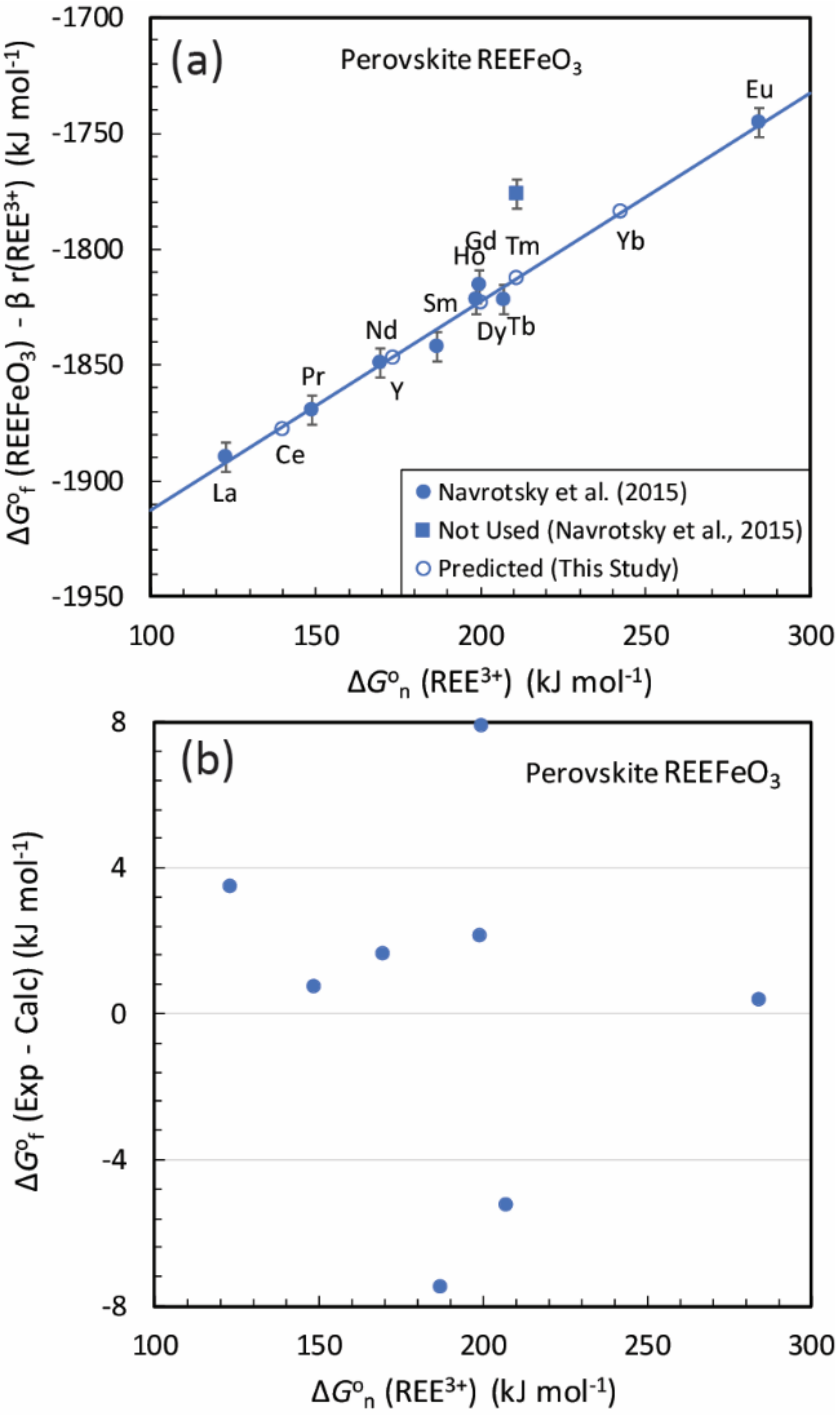

**Figure 9. (a)** Graphical representations of linear correlations of isostructural perovskite $REEFeO_3$ mineral families, and **(b)** Residuals between the experimentally measured and calculated $\Delta G^o_f$ values for perovskite $REEFeO_3$ isostructural family.

Figure 10

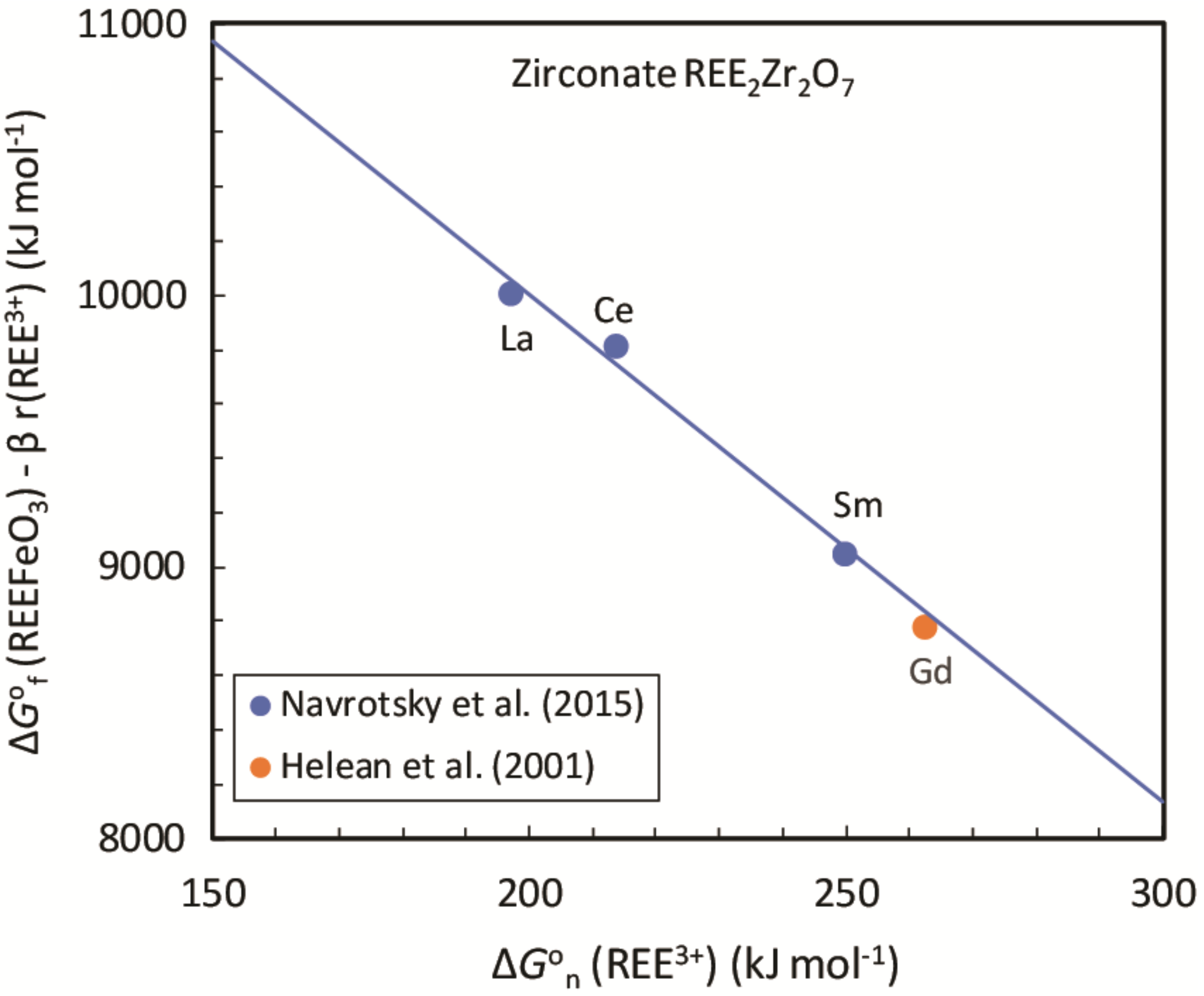

**Figure 10.** Graphical representations of linear correlations of isostructural zirconate $REE_2Zr_2O_7$ mineral families. Only the $Gd_2Zr_2O_7$ data is from Helean et al. (2000) to construct the linear correlation. The negative slope indicates that the selected dataset for zirconate is not recommended.

Figure 11

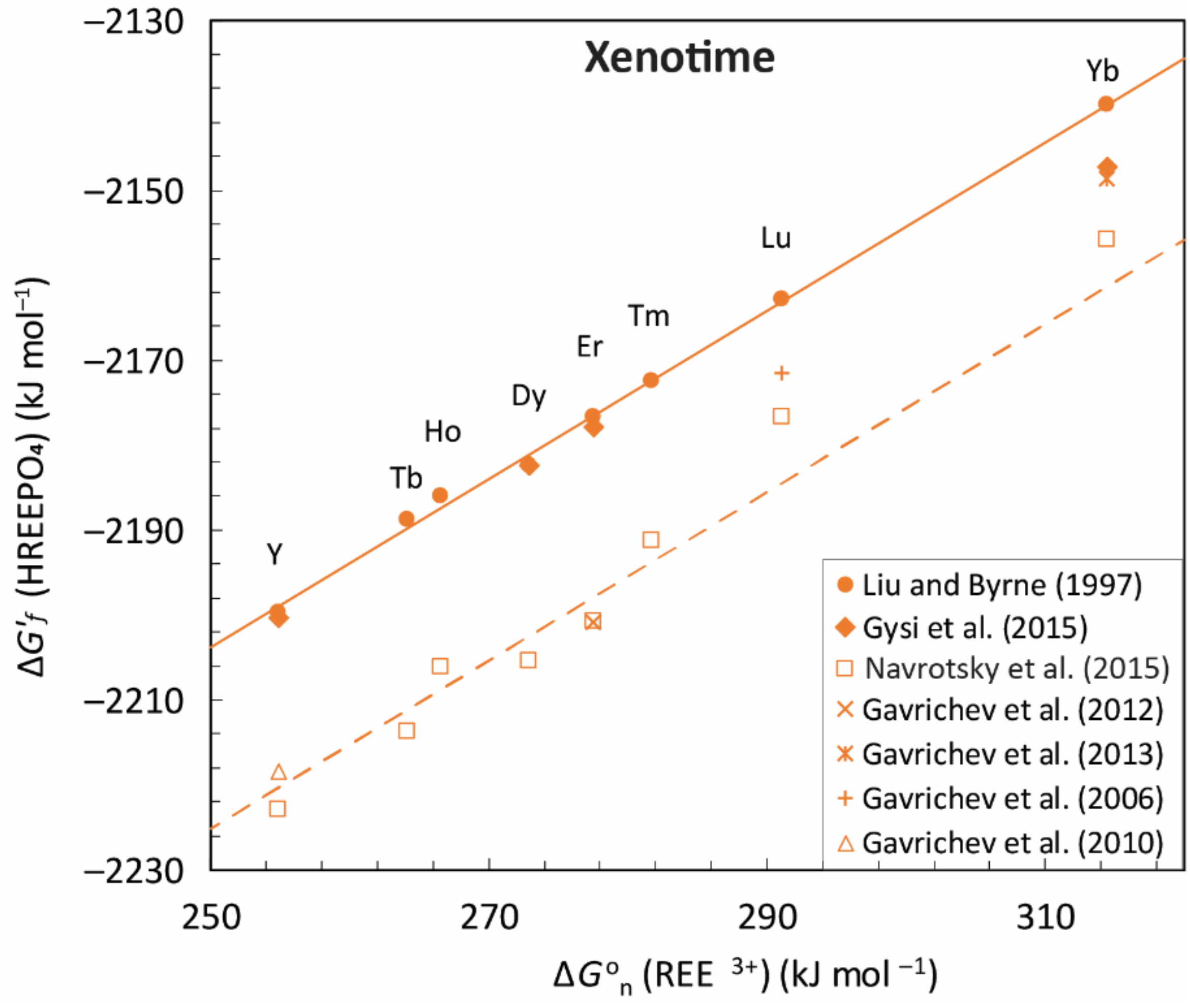

**Figure 11.** Graphical representations of linear correlations of isostructural xenotime and comparisons with other experimental studies (Gavrichev et al., 2013; Gavrichev et al., 2012; Gavrichev et al., 2010; Gavrichev et al., 2006; Gysi et al., 2015; Liu and Byrne, 1997; Navrotsky et al., 2015). Solid symbols represent mineral solubility experiments and open symbols represent calorimetric experiments.

Figure 12

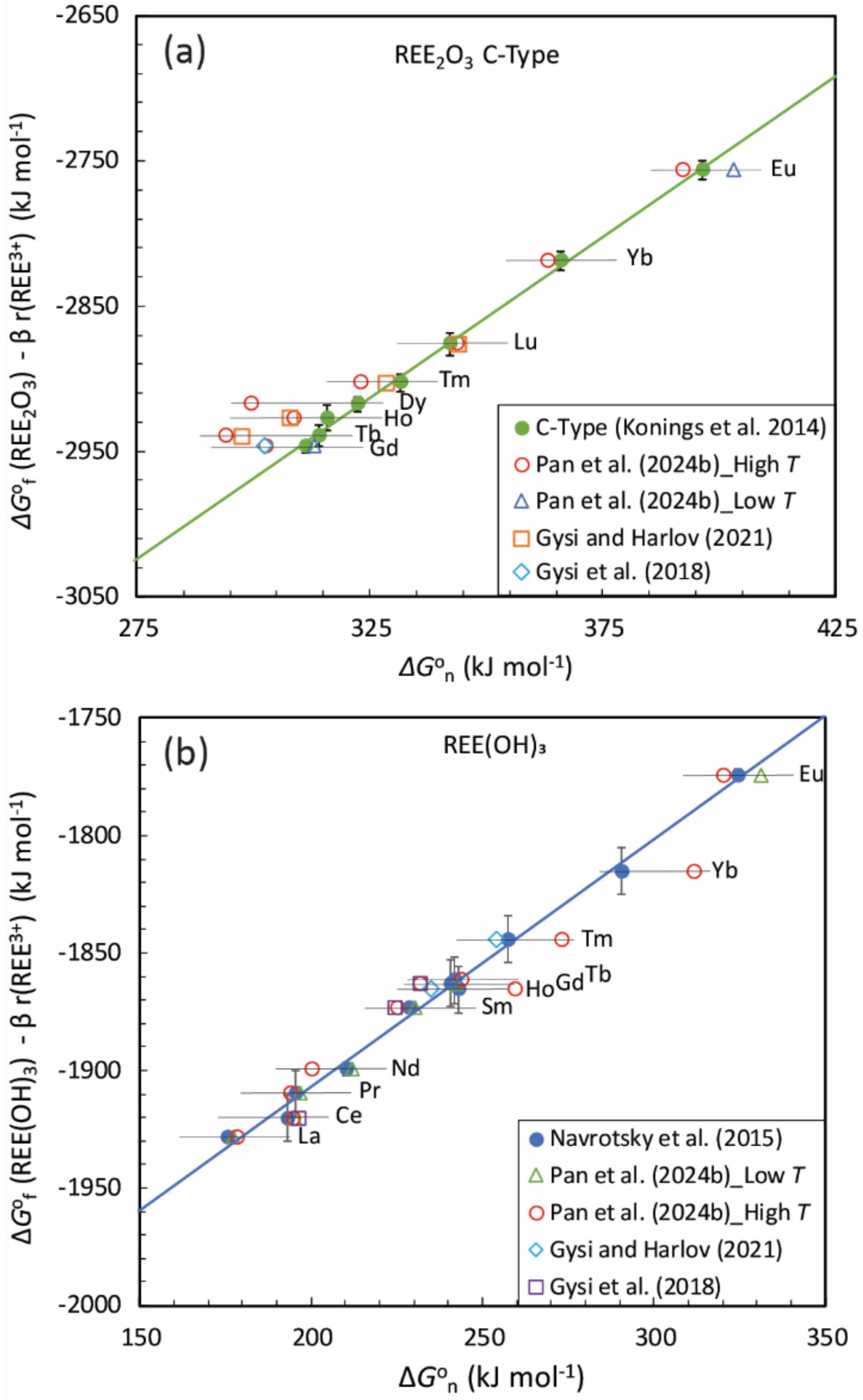

**Figure 12.** Linear correlations of **(a)** $REE_2O_3$ C-Type and **(b)** REE hydroxides constructed with the $REE^{3+}$ from Shock et al. (1997) and comparisons with previously reported $\Delta G^o{}_f$ values in literature (Gysi and Harlov, 2021; Gysi et al., 2018; Konings et al., 2014; Pan et al., 2024b).

Click here to access/download

**Supplementary Material**

Appendix A.pdf